\documentclass[prl, twocolumn,  floatfix, superscriptaddress, aps, amsmath, amsfonts]{revtex4-1}  
\usepackage{graphicx} 
\usepackage[english]{babel}

\usepackage{array}
\usepackage[T1]{fontenc}
\usepackage[usenames,dvipsnames]{xcolor}

\usepackage{mathptmx}
\usepackage{bm}
\usepackage{xcolor}

\newcommand{\add}[1]{#1} 
\newcommand{\rem}[1]{\textcolor{red!50!gray}{}}

\definecolor{cream}{RGB}{222,217,201}

\begin{document}

\title{Acoustic bubble dynamics in a yield-stress fluid}

\author{Brice Saint-Michel}
\altaffiliation{Present address: Department of Chemical Engineering, Delft University of Technology, Delft 2629 HZ, the Netherlands}
\affiliation{Department of Chemical Engineering, Imperial College London, London SW7 2AZ, United Kingdom} 

\author{Valeria Garbin}
\email[]{v.garbin@tudelft.nl}
\altaffiliation{Present address: Department of Chemical Engineering, Delft University of Technology, Delft 2629 HZ, the Netherlands}
\affiliation{Department of Chemical Engineering, Imperial College London, London SW7 2AZ, United Kingdom}

\begin{abstract}
Yield-stress fluids naturally trap small bubbles when their buoyancy applies an insufficient stress to induce local yielding of the material. Under acoustic excitation, trapped bubbles can be driven into volumetric oscillations and apply an additional local strain and stress that can trigger yielding and assist their release. In this paper we explore different regimes of microbubble oscillation and translation driven by an ultrasound field in a model yield-stress fluid, a Carbopol microgel. We first analyse the linear bubble oscillation dynamics to measure the local, high-frequency viscosity of the material. 
We then use acoustic pressure gradients to induce bubble translation and examine the elastic part of the response of the material below yielding. We find that, at moderate pressure amplitude, the additional stresses applied by volumetric oscillations and acoustic radiation forces do not lead to any detectable irreversible bubble motion.  At high pressure amplitude, we observe non-spherical shape oscillations that result in erratic bubble motion. The shape modes and critical pressures we observe differ from the predictions of a recent model of shape oscillations in soft solids.
Based on our findings, \rem{we provide guidelines on how to use an acoustic field to optimise bubble removal in yield-stress fluids}
\add{we discuss possible reasons for the lack of bubble release in Carbopol and suggest other systems in which ultrasound-assisted bubble rise may be observed.}

\end{abstract}

 \maketitle

\section{Introduction}

Yield-stress fluids encompass a wide range of materials including foams, suspensions, emulsions and microgels~\cite{Coussot2014,Bonn2017}. These materials exhibit a threshold in applied stress, called the yield stress, below which the material behaves like a solid, and above which it flows like a liquid. A clear manifestation of the yield stress is the presence of trapped bubbles, when their buoyancy force is too small to yield the material. Trapped bubbles can be beneficial, for instance when they are used to impart texture to a food product, or they can be detrimental as they can negatively affect the thermal conductivity or optical transparency of a material. 
Strategies to control the amount and size distribution of trapped bubbles are therefore important in processing of formulations and advanced materials. There is some experimental evidence that driving bubbles into volumetric oscillations in yield-stress fluids can assist their removal~\cite{Stein2000}, but the effect of oscillations on yielding is poorly understood. This lack of understanding is particularly detrimental to the development of controlled bubble removal methods.

Understanding  bubble dynamics in yield-stress fluids is particularly challenging since their rheology is not even fully understood in the case of simple shear. Yield-stress fluids are only well-understood in the limit of very small shear stresses a linear elastic behaviour is recovered, or for large, steady  stresses for which their flow rheology usually obeys the Herschel-Bulkley equation~\cite{Coussot2014}. For intermediate stresses, experimental results performed under steady or large amplitude oscillatory shear~\cite{Hyun2011} have evidenced that yield-stress fluids exhibit non-linear~\cite{Lidon2017a}, time-dependent, cooperative~\cite{Goyon2008} behaviour. Such features are only captured by the most recent microscopic~\cite{Nicolas2018} and continuum mechanics models~\cite{Dimitriou2019}.

The capacity of a yield-stress fluid to entrap bubbles up to a critical effective radius $R_{\rm c}$ 
can be expressed as the dimensionless number $Y_{\rm c}^{-1} = 2 \rho_{\rm l} g R_{\rm c} / 3 \sigma_{\rm Y}$, where $\rho_{\rm l}$ is the liquid density, $g$ is the acceleration due to gravity and $\sigma_{\rm Y}$ is the yield stress of the material~\cite{Dimakopoulos2013}. 
Even with the most conservative estimate, a yield stress of only $10$~Pa causes trapping of bubbles up to $2R_{\rm c} = 6$~mm in diameter. Removing bubbles below the critical size can be achieved by centrifuging \cite{Mazzeo2012}, applying a vacuum,  or by using low-frequency ($\sim 100$~Hz) vibrations to suppress the yield stress in fragile granular networks~\cite{Koch2019}. These techniques alter the physical parameters at play in the definition of $Y_{\rm c}^{-1}$ rather than fundamentally altering this criterion.

Bubbles are however not passive under the application of vibrations and acoustic excitations, as the dynamic pressure field drives them into volumetric oscillations~\cite{Plesset1977}. Oscillating bubbles apply a local strain field to the surrounding material, which in turn reacts by exerting a stress onto the bubble, altering the oscillation dynamics. Bubble dynamics in Newtonian liquids~\cite{Plesset1977} and soft solids~\cite{Dollet2019} is now a well-established topic, motivated e.g. by the direct role played by bubble collapse in therapeutic laser or ultrasound tissue ablation~\cite{Coussios2008,Barney2020}. Bubble radius time profiles are now even used either in the linear regime~\cite{Jamburidze2017} or the strongly non-linear, cavitation regime~\cite{Estrada2018} to extract local rheological properties of soft solids.

In yield-stress fluids, oscillating bubbles may apply a strain that is sufficient to locally yield the material, defining a yielded, fluid region. Bubble rise may then proceed in this confined region even if their size is well below $R_{\rm c}$. The size and shape of this region has a key influence on the bubble rising velocity, and ultimately in the efficiency of the removal process. While the shape of the yielded region has been investigated in great detail for passive bubble rise~\cite{Holenberg2013,Dimakopoulos2013}, the case of oscillating bubbles has only been examined very recently~\cite{Karapetsas2019,DeCorato2019}. Building upon the progress in modelling both bubble dynamics in soft materials~\cite{Dollet2019} and the rheology of yield-stress fluids~\cite{Saramito2009}, these articles confirm that bubble rise is indeed possible for bubbles below $R_{\rm c}$~\cite{Karapetsas2019}; they also compute the minimum oscillation amplitude required to initiate yielding~\cite{DeCorato2019}. To the best of our knowledge, these numerical and theoretical results have not yet been compared to experiments: experimental articles so far have focused on the case of bubble removal in a shear-thinning, viscoelastic surrounding fluid~\cite{Iwata2008} and removal in yield-stress fluids for bubbles already close to the static rise radius at rest, $R_{\rm c}$~\cite{Stein2000}.

In this article, we conduct experiments to test the criterion for medium yielding and bubble removal that we previously derived~\cite{DeCorato2019}, 
using a Carbopol microgel as a model yield-stress fluid. We investigate the oscillation dynamics of initially spherical bubbles ($100-200~\mu$m) excited by a standing-wave ultrasound field with controlled frequency ($19-30$~kHz), acoustic pressure amplitude, and spatial distribution of pressure gradients. We measure the resonance curve of the bubbles, their mobility in a pressure gradient and the onset of non-spherical shape oscillations. We extract the viscosity and linear elastic modulus of the material, and compare these measurements to the predictions of the model \cite{DeCorato2019}. We finally conclude on the efficiency of bubble removal through bubble oscillation in yield-stress fluids.


\section{Bubble dynamics in yield-stress fluids}
\label{sec:theory}

\subsection{Governing equations for bubble oscillations}
\label{sec:theory_equationsbubbleosc}

We briefly recall here the physics of the linear oscillations of a spherical bubble in a yield-stress fluid we derived in a previous article \cite{DeCorato2019}. We will show in Section~\ref{sec:experiment_dimensionless} that the assumptions of spherical bubble and linear dynamics are reasonable given the size of the bubbles and the rheological properties of the fluid that we use in the experiments. 

A bubble with equilibrium radius $R_0$ is driven into volumetric oscillations under an acoustic excitation at a frequency $f$, i.e. a sinusoidal applied pressure $p(t) = p \sin (2 \pi f t)$ far away from the bubble. The time-dependent radius, $R(t)$, is:
\begin{equation}
    R(t) = R_0 \left [1 +  \zeta(t) \right ]\,.
\end{equation}
Applying the momentum and the mass conservation for the fluid between the spherical bubble surface $r = R(t)$ and $r \to \infty$ yields a generalised Rayleigh-Plesset equation valid for arbitrary fluids~\cite{Prosperetti1982}. \add{Previously our group has derived a model for bubble dynamics in yield-stress fluids by combining the generalised Rayleigh-Plesset equation~\cite{Prosperetti1982} with the elasto-visco-plastic rheological model proposed by~\citet{Saramito2009}. The details of the full model can be found in Ref.~\citenum{DeCorato2019}. We recall here that for small-amplitude oscillations and below the yield point, the rheological model reduces to a Kelvin-Voigt viscoelastic solid of linear elastic modulus $G$ and solvent viscosity $\eta_{\rm s}$. A Taylor expansion of the momentum balance valid at order 1 in $\zeta$ may then be derived following the classical linear theory of bubble dynamics~\cite{Prosperetti1977}:}
\begin{equation}
    \label{eq:raylpless_linear}
    \ddot\zeta + 2 \beta \dot\zeta + 4 \pi^2 f_{\rm 0}^2 \zeta = - \frac{p}{\rho R_0^2} \sin (2 \pi f t)\,,
\end{equation}
in which $\rho$ is the fluid density, assumed to be a constant, and $\beta$ and $f_{\rm 0}$ are respectively the damping coefficient and the natural frequency of the bubble oscillations. These two quantities depend \emph{a priori} on the rheology of the fluid. 

Equation~\eqref{eq:raylpless_linear} is a standard second-order linear differential equation that we can reformulate in the frequency domain. We then obtain the second-order transfer function for the bubble oscillation amplitude $\zeta$ in the spirit of earlier works on bubble spectroscopy~\cite{vanderMeer2007,Hamaguchi2015,Jamburidze2017}:

\begin{subequations}
\begin{align}
    \zeta(t) &= \zeta \sin(2 \pi f t + \phi) \\
    \label{eq:resonance_curve}
    \zeta &= \frac{p / \rho R_0^2}{4 \pi \sqrt{ \pi^2 \left ( f_{\rm 0}^2 - f^2 \right )^2 + \beta^2 f^2}} \\
    \label{eq:phase_lag}
    \phi &= \frac{\pi}{2} + \arctan \left [ \frac{\pi}{\beta f} (f_{\rm 0}^2 - f^2) \right ]
\end{align}
\end{subequations}
The amplitude part of the transfer function [Equation~\eqref{eq:resonance_curve}] gives the resonance curve of the bubble. The phase lag between the bubble oscillation and the pressure field $\phi$ made explicit in Equation~\eqref{eq:phase_lag} spans from $\pi$ for $f \ll f_0$ in the low frequency case to $0$ for $f \gg f_0$ in the high frequency case. 

The natural oscillation frequency $f_0$ based on the model of~\citet{Saramito2009} is derived in Ref.~\cite{DeCorato2019}:
\begin{equation}
    \label{eq:minnaert_full}
    f_{\rm 0}^2 = \frac{3 \kappa p_0 + 2 (3 \kappa - 1 ) \Gamma /R_0 + 4 G}{4 \pi^2 \rho R_0^2}\,,
\end{equation}
in which $\Gamma$ is the surface tension between the gas and the fluid \add{and $p_0$ is the ambient atmospheric pressure} \rem{and $G$ is the linear elastic modulus of the yield-stress fluid}. We also introduce here the polytropic exponent $1.0 \leq \kappa \leq 1.4$ that indicates the nature of the thermodynamic process occurring in the bubble, from isothermal ($\kappa = 1$) to adiabatic ($\kappa = 1.4$) depending on the thermal Péclet number~\cite{Prosperetti1977}.

For very soft materials for which $G \ll p_0$, and for sufficiently large bubbles, i.e. for $R_0 \gg \Gamma / p_0 = 1.0~\mu$m, we recover the standard Minnaert frequency~\cite{Minnaert1933} for a given bubble radius $R_0$:
\begin{equation}
    \label{eq:minnaert_simple}
   f_{\rm m} = \frac{1}{2 \pi R_0} \sqrt{\frac{3 \kappa p_0}{\rho}}\,.
\end{equation}
Equation~\eqref{eq:minnaert_simple} may be used as well to derive a resonant radius $R_{\rm m}$ for a given oscillation frequency $f$. We also recall the predictions for the damping parameter $\beta$~\cite{Hamaguchi2015,DeCorato2019}:
\begin{equation}
    \label{eq:damping}
        \beta = \frac{2 \eta_{\rm eff}}{\rho R_0^2} = \frac{2}{\rho R_0^2} \left [ \eta_{\rm s} + \frac{\pi^2 \rho f^2 R_0^3}{c} + \frac{3 p_0 \kappa'}{8 \pi f}  \right ]
\end{equation}
The three terms at the right hand side of Equation~\eqref{eq:damping} respectively account for viscous dissipation \add{proportional to the solvent viscosity $\eta_{\rm s}$ in the Kelvin-Voigt model}; acoustic scattering of the bubble, and thermal dissipation, in which the dimensionless quantity $\kappa'$ is related to the polytropic exponent $\kappa$ introduced earlier~\cite{Prosperetti1977}. Appendix~\ref{sec:app:damping} shows the relative magnitude of each contribution to $\beta$ for our experiments. \add{The relative uncertainty on these quantities is discussed in ESI Section 1}. 

\subsection{Acoustic radiation forces}
\label{sec:theory_acousticradiation}

Gradients in a pressure field exert a force ${\bf F} = - V {\bm \nabla} \! p$ on objects of volume $V$. In a standing wave field $p({\bf x}, t) = p({\bf x}) \sin (2 \pi f t)$, the average force $\langle {\bf F} \rangle$ applied on an incompressible object of fixed volume $V$ over one oscillation cycle is zero. Because bubbles expand and contract in response to oscillations in pressure, the same pressure gradient applies a larger net force on the object when its radius is large than when it is small. This leads to a net force over one oscillation period called Bjerknes force~\cite{Crum1975}:

\begin{align}
    \label{eq:bjerknes}
    \langle{\bf F} ({\bf x})\rangle &= -2 \pi R_0^3 {\bm \nabla} \! p ({\bf x}) \zeta \cos(\phi) \,.
\end{align}
For a driving frequency $f$ and an equilibrium bubble size $R_0$, small bubbles for which $\cos(\phi) = -1$ will move towards high pressure areas (named anti-nodes) whereas large bubbles for which $\cos(\phi) = +1$ will move towards low pressure areas (nodes), a classical result in Newtonian fluids~\cite{Leighton1990}. Bjerknes forces are non-linear as both ${\bm \nabla \! p}$ and $\zeta$ are proportional to the applied pressure. They are particularly efficient at pushing and pulling bubbles against gravity when the relative pressure gradient $|{\bm \nabla} \! p / p|$ is high. 

Following Equation~\eqref{eq:resonance_curve} the pressure $p$ required to obtain a constant oscillation amplitude $\zeta$ for all bubble radii $R_0$ is much higher far away from the resonance condition than at resonance. As a consequence, for an imposed oscillation amplitude $\zeta$ the pressure gradient ${\bm \nabla} \! p$ in Equation~\eqref{eq:bjerknes} and the Bjerknes forces will also be stronger away from resonance. We will use this strategy in Section~\ref{sec:results_motion} to apply strong Bjerknes forces while remaining in the linear range of the bubble oscillation amplitude $\zeta$.

Recent articles have related the force applied to spherical objects and their displacement in purely elastic~\cite{Ilinskii2005} or \add{Kelvin-Voigt} viscoelastic solids~\cite{Urban2011}, which can then be applied to yield-stress fluids for relatively small deformations. Assuming the pressure gradient ${\bm \nabla} \! p$ at location ${\bf x}$ is directed alongside $z$ we have:
\begin{equation}
    \label{eq:displacement_linelastic}
    \frac{\Delta z ({\bf x})}{R_0} = \frac{1}{R_0} \left \langle\frac{F_z ({\bf x},t)}{4 \pi G R(t) } \right \rangle = - \frac{1}{3} \frac{ R_0 }{G} {\bf \nabla\!}_z p ({\bf x}) \zeta \cos(\phi)
\end{equation}
\add{Equation~\eqref{eq:displacement_linelastic} remains valid as long as the oscillations do not alter the properties of the fluid. Interestingly, it provides a measurement of $G$ that is unaffected by $\Gamma$ and $p_0$ in contrast with Equation~\eqref{eq:minnaert_full}. We will use Equation~\ref{eq:displacement_linelastic} to measure $G$ in Section~\ref{sec:results_motion}}.

\rem{This prediction supposes that the bubble oscillations do not modify the properties of the fluid. We will use it in Section IV C to compute the elastic modulus $G$ of the fluid using the motion of the centre of the bubble}

\subsection{Yielding criteria \add{and impact on bubble dynamics}}
\label{sec:theory_yieldingcriterion}

\paragraph*{Yielding to oscillations}
When no pressure gradient is present, the centre of the bubble is not moving and the strain field is spherically symmetric. Its expression in the spherical reference frame ($r, \theta, \varphi$) centred on the bubble reads~\cite{Macosko1994}:

\begin{equation}
    \epsilon_{rr} (r,t) = \left (1 + \frac{R(t)^3 - R_0^3}{r^3} \right)^{-4/3}- 1 \simeq - 4 \zeta(t) \frac{R_0^3}{r^3}
    \label{eq:strain_profile}
\end{equation}

\rem{This strain can be related to the total stress in the unyielded fluid. In the framework of Saramito, the total stress includes a viscous contribution from the solvent -–the Newtonian part of the stress– equal to 2 $\eta_{\rm s} {\bm \dot\epsilon}$ and an elastic contribution $2 G{\bm \epsilon}$ -- the non-Newtonian part of the stress -- below the yield stress.}
\add{The Kelvin-Voigt model, assumed to be valid below yielding, expresses the applied stress as a sum of an elastic stress $ G{\bm \epsilon}$ and a viscous stress $\eta_{\rm s} \dot{\bm \epsilon}$. For sufficiently large oscillation amplitudes, the elastic stresses may satisfy the von Mises yield criterion~\cite{Hill1998,DeCorato2019} in a corona of fluid surrounding the bubble. The material then follows a Kelvin-Voigt rheology only outside of the yielded region, including at its edge, located at a distance $r_{\rm Y}$ from the centre of the bubble:}
\begin{equation}
     \label{eq:yielding_vonMises}
     \left (\frac{r_{\rm Y}}{R_0} \right )^3 = \frac{2 \sqrt{3} G}{\sigma_{\rm Y}} |\zeta(t)| 
\end{equation}
Equation~\eqref{eq:yielding_vonMises} defines the extent \add{$r_{\rm Y}$} of the yielded region as a function of time. Fluid yielding starts when the yielded region exceeds the bubble size at rest $R_0$ at least once during an oscillation cycle. This simplified yielding criterion reads $\zeta \geq \zeta_{\rm c} = \sigma_{\rm Y}/ 2 \sqrt{3} G$ and we hypothesise it is a necessary condition to initiate irreversible bubble rise.

\add{In the yielded region, the purely elastic component of the Kelvin-Voigt model becomes a Maxwell element~\cite{Saramito2009}, keeping its elastic modulus $G$ and adding a non-linear plastic degree of deformation of viscosity $\eta_{\rm evp}$, traditionally defined as $K \dot\epsilon^{n-1}$ in rotational rheology. The elasto-plastic crossover time of the yielded material is $(K/G)^{1/n}$: the yielded material remains predominantly elastic below this time scale while plastic deformation dominates above it. Bubble oscillation dynamics is then only affected by yielding when the applied frequency satisfies $2 \pi f (K / G)^{1/n} \leq 1$, in agreement with numerical simulations~\cite{DeCorato2019}.}

\add{Bubbles also apply a constant stress onto the fluid due to buoyancy or acoustic radiation forces. Hence, these forces will act on the yielded material during the whole time $N/f$ of the acoustic excitation. Irreversible bubble displacement may then be observed provided that $f/N (K/G)^{1/n} \leq 1$.}

\paragraph*{Yielding to acoustic radiation forces}
A second bubble release criterion can be computed from acoustic radiation forces, \add{ignoring the contribution of the oscillatory stresses}. We can compare the average acoustic radiation stress $\sigma_{\rm ac} = \langle{F_z / 2 \pi R^2}\rangle$ to the yield-stress in direct analogy with the yielding parameter $Y_c^{-1}$ used for gravity-driven bubble rise. This critical parameter varies between $1.1$ for the most efficient, inverted teardrop shapes~\cite{Sikorski2009} to $5.1$ for bubbles that are almost spherical~\cite{Dimakopoulos2013}, which we consider in this article. Acoustic radiation forces then initiate bubble rise provided that:
\begin{equation}
    \label{eq:yielding_acousticradiation}
    \frac{1}{\sigma_{\rm Y}} \underbrace{\left \langle \frac{F_z ({\bf x},t)}{2 \pi R^2(t) } \right \rangle}_{\sigma_{\rm ac}}  = \frac{1}{3} \frac{  R_0 |{\bm \nabla} \! p ({\bf x})| }{\sigma_{\rm Y}} \zeta |\cos( \phi)|  \geq 5.1\,.
\end{equation}


\section{Materials and Methods}
\label{sec:experiment}

\subsection{Carbopol microgel preparation and properties}
\label{sec:experiment_carbopol}

The yield-stress fluid we use in this article is a Carbopol ETD 2050 microgel (Lubrizol Corporation, Wickliffe, Ohio, U.S.A.) of concentration~0.15\% w/v that has been extensively studied in the literature~\cite{Piau2007,Lefrancois2015,Lidon2017a,Dinkgreve2018}. The Carbopol primary particles are made of crosslinked polyacrylic acid, which swells at high pH to form a jammed assembly of soft particles with a diameter of several microns~\cite{Lefrancois2015}.

Following classical preparation protocols~\cite{Lidon2017a,Dinkgreve2018}, we first let the Carbopol flakes dissolve in MilliQ water (18.2 M${\Omega}$.cm) for 1~hour under gentle agitation before adding 1\% v/v 1M NaOH to adjust the pH to 7. The fluid is then stirred for 20 minutes by an overhead mixer (RW 20 fitted with a R1303 dissolver impeller, IKA, Staufen im Breisgau, Germany) at 2000 rpm. We then place the fluid in a vacuum chamber until all bubbles that have been incorporated during mixing are removed. The fluid is finally left to equilibrate overnight.

We characterise the rheology of the Carbopol microgel using a rotational rheometer (MCR 302, Anton Paar, Graz, Austria). We perform flow curves and oscillatory measurements, from which we deduce $\sigma_{\rm Y}$ and $G$ following standard fits~\cite{Coussot2014}; both data series are displayed in Appendix~\ref{sec:app:rheology}. We measure the sound velocity in the fluid $c$ using a separate acoustic setup. We assume that its density is equal to that of water at room temperature and we choose a surface tension $\Gamma$ based on dedicated experiments eliminating the impact of elastic stresses~\cite{Jorgensen2015}. We finally use the standard heat diffusivity $D$ of air from classical sources~\cite{CRC2019} to compute the thermal dissipation coefficient $\kappa'$ from Section~\ref{sec:theory_equationsbubbleosc}. The values of these parameters are compiled in Table~\ref{tab:parameters}.

\begin{table}
    \centering
    \caption{Physical parameters of bubble oscillation in Carbopol. Source of the data: $n$, $K$, $\sigma_{\rm Y}$, $G$ and $c$ have been measured by the authors. The surface tension $\Gamma$ and the heat diffusivity $D$ and taken from Refs. \citenum{Jorgensen2015} and \citenum{CRC2019} respectively. The polytropic index $\kappa$ is computed following Ref.~\citenum{Prosperetti1977}.}
    \begin{tabular}{rccrl}
        \hline
        Name                & Fluid     & Symbol            & Value   & Unit \vspace*{0.5em} \\
        Polytropic Index    &           & $\kappa$          & $1.30$          &  \\
        Ambient Pressure    & Air       & $p_0$             & $1.013~10^5$    & Pa \\
        Heat diffusivity    & Air       & $D$               & $1.9~10^{-5}$   & m$^2$.s$^{-1}$\\
        Viscosity           & Water     & $\eta_{\rm 0}$    & $1.0~10^{-3}$   & Pa.s \\
        Specific gravity    & Water     & $\rho$            & $9.98~10^{2}$   & kg.m$^{-3}$\\
        Sound velocity      & Carbopol  & $c$               & $1.495~10^{3}$  & m.s$^{-1}$\\
        Surface Tension     & Carbopol  & $\Gamma$ & $6.2~10^{-2}$   & N.m$^{-1}$\\
        Flow Index          & Carbopol  & $n$               & $0.36$          & \\
        Flow Consistency    & Carbopol  & $K$               & $5.0$           & Pa.s$^n$\\
        Yield Stress        & Carbopol  & $\sigma_{\rm Y}$  & $5.3$           & Pa \\
        Shear Modulus       & Carbopol  & $G$               & $36.0$          & Pa \\
    \end{tabular}
    \label{tab:parameters}
\end{table}

\subsection{Ultrasound excitation and high-speed imaging}
\label{sec:experiment_acoustic_setup}

\begin{figure}
    \centering
    \includegraphics[width=8cm]{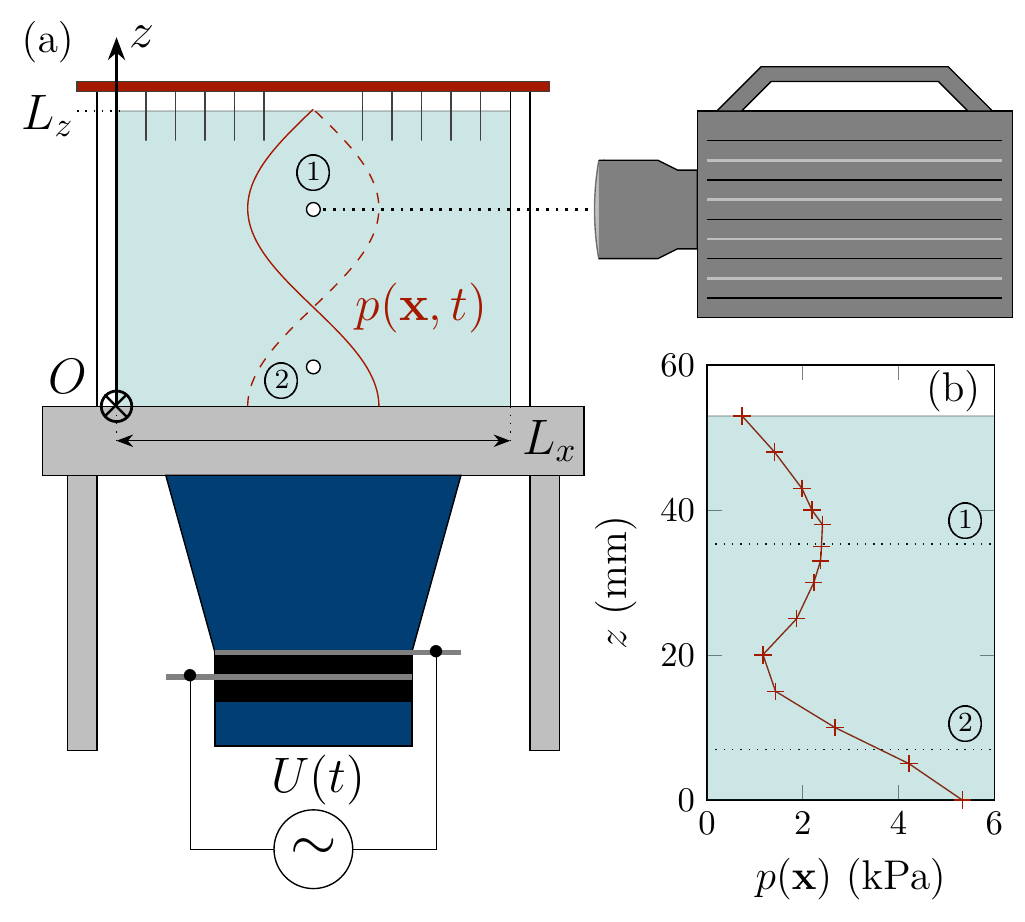}
    \caption{(a) Schematic diagram of the experiment. A Langevin transducer (in blue) provides an acoustic excitation to the yield-stress fluid container above it. The excitation frequency $f$ matches the resonance of the transducer and that of a $(0,0,3/2)$ standing wave pattern $p({\bf x}, t)$ inside the container. Bubbles of interest are illuminated by an optical fibre facing the high-speed camera (right, in grey). (b) Vertical pressure profile in water at the centre of the setup for a frequency $f = 21.5$~kHz, an applied voltage $U = 5.0$~V and a fluid height $L_z = 5.3$~cm. We report the vertical positions of  locations \textcircled{\footnotesize 1} and \textcircled{\footnotesize 2} (dashed lines).}
    \label{fig:setup}
\end{figure}

Our experiments take place in a parallelepipedic container filled with the yield-stress fluid, as sketched in Figure~\ref{fig:setup}. The walls of the containers are either made of glass or duralumin, ensuring total internal reflection of the incident acoustic wave. A lid fitted with needles partially dipped in the fluid is used at the top of the device to prevent sloshing while maintaining the total internal reflection with air.

We apply acoustic excitations using a Langevin transducer (Steminc, Doral, Florida, U.S.A.) oscillating between $f = 19.45$ and $29.2$~kHz.  We drive the transducer using a waveform generator (33210A, Agilent, Santa Clara, U.S.A.) coupled to a linear amplifier (AG 1021, T\&C Power Conversion, Rochester, U.S.A.). The amplifier gain controls the voltage $U$ applied to the transducer and ultimately the applied pressure amplitude $p({\bf x},t)$ during the experiment. We always work at relatively low input voltage and amplifier gain to prevent non-linear distortion of the amplifier or transducer response. 

The container dimensions $L_x = 10.2$~cm, $L_y = 5$~cm, and $L_z$ are adapted to produce a resonant standing wave pattern at the applied frequency $f$, where the pressure amplitude $p({\bf x})$ varies mostly alongside ${\bf e_z}$. This pattern, shown in Figure~\ref{fig:setup}(a), corresponds to the $(0,0,3/2)$ room mode of the container~\cite{Morse1944}. Pressure measurements using a polyvinylidene fluoride hydrophone (RP 42s, RP Acoustics, Leutenbach, Germany) along the vertical line at the centre of the container [presented in Figure~\ref{fig:setup}(b)] are compatible with the predicted room mode; they also show that the distortion level is small. We then define two locations named \textcircled{\footnotesize 1} and \textcircled{\footnotesize 2} (see Figure~\ref{fig:setup}). The first location corresponds to the pressure anti-node at two-thirds of the cell height for which the  pressure gradient ${\bm \nabla} \! p({\bf x}_1)$ is zero. It is used in Sections~\ref{sec:results_linear} \add{and} \ref{sec:results_resonancecurve} \rem{and \ref{sec:results_shapeinstab}}. The second location is chosen below the pressure node to achieve both a significant pressure and pressure gradient so as to maximise acoustic radiation forces, as explained in Section~\ref{sec:theory_acousticradiation}. At this location and for an applied frequency $f = 19.45$~kHz used throughout Section~\ref{sec:results_motion}, we measure a  relative pressure gradient $| {\bm \nabla} \! p ({\bf x}_2) / p ({\bf x}_2) |$ in the vertical direction equal to $82$~m$^{-1}$. Given the efficiency of the resonant setup $p({\bf x}_2)/U = 0.13$~kPa.V$^{-1}$ at this location, the acoustic pressure gradient $|{\bm \nabla} \! p ({\bf x}_2)|$ exceeds the hydrostatic pressure gradient for voltages $U \geq 1$~V.

We align the high-speed camera (Fastcam SA5, Photron, Tokyo, Japan) at location \textcircled{\footnotesize 1} or \textcircled{\footnotesize 2} by imaging the tip of the hydrophone. We then remove the hydrophone and inject a bubble of initial radius $\leq 300~\mu$m with a small syringe in the frame of the camera before fine-tuning its position through careful manual pushing. Such a procedure inevitably modifies the internal stresses of the fluid around the bubble. Following Ref.~\citenum{Hamaguchi2015}, we consider the slow bubble dissolution (reported in ESI Section 2) as a sweep over the initial bubble radii $R_0$ and we then produce a resonance curve [Equation~\eqref{eq:resonance_curve}] for a constant frequency $f$ and varying $R_0$.
At the start of the camera acquisition, a burst of $N = 200$ to $3000$ sinusoidal cycles is sent by the waveform generator to the amplifier and the transducer. The camera records images up to $250\,000$ frames per second, corresponding to $\sim 10$ images per oscillation period. We set the total acquisition time to measure both the bubble response to the acoustic excitation and its subsequent relaxation. We report in our acquisitions a small source of vibration at $f = 130$~Hz. It impacts bubble position measurements but does not affect the measured bubble radius and shape.

\subsection{Bubble contour detection and decomposition into Legendre and Fourier modes}
\label{sec:experiment_dataprocessing}

The data processing of video acquisitions is inspired by two recent works~\cite{Jamburidze2017,Guedra2017}. In short, our {\scshape Matlab} routines normalise raw images and apply a luminosity threshold in order to retrieve the location of the bubble centroid ${\bf x}(t)$ and its mean radius $R(t) = R_0 [1 + \zeta(t)]$ as a function of time.

We then perform a Fourier transform of the radius time series and pay a particular attention to $\hat \zeta(\nu)$ when $\nu$ is a multiple or a sub-multiple of the oscillation frequency $f$. Significant harmonic content indicates that we are no longer working in the linear bubble oscillation framework described in Section~\ref{sec:theory}.

\add{We also study whether bubbles remain spherical during the oscillations by examining the two-dimensional outline of the bubbles. To do so, we plot $360$ lines originating at the bubble centroid, with equally spaced polar angles $\theta$, defined from the vertical direction as shown in Figure~\ref{fig:analysis_linear}(a). We define the local bubble radius $R_0 [1 + \zeta(\theta,t)]$ as the point where each line crosses the bubble edge. We then define the bubble orientation $\theta_0(t)$ as the angle for which $R(\theta_0+\theta,t)$ lies closest to $R(\theta_0-\theta,t)$. In earlier studies~\cite{Versluis2010,Hamaguchi2015,Poulichet2017,Guedra2017}, bubble outlines usually show a clear $k$-fold symmetry, which is empirically assumed to correspond to the degree, or mode, $k$ of the spherical harmonics $Y_k^m$ describing the three dimensional shape of the bubble. Following the same approach, we project the bubble shape outline $\zeta(\theta,t)$ on the Legendre polynomials of degree $k$, $P_k (\cos(\theta))$~\cite{Guedra2017}. We may then define the instantaneous amplitude of a shape mode $k$, $\zeta_k(t)$:}

\rem{We also measure the shape of the bubble by plotting 360 lines originating at the bubble centroid, with equally spaced polar angles $\theta$, defined from the vertical direction as shown in Figure 2(a).  We define the local bubble radius $R_0 [1+\zeta(\theta,t)]$ as the point where each line crosses the bubble edge. We then project the bubble shape function $\zeta( \theta,t)$ on the Legendre polynomials of degree $k$,$P_k(\cos(\theta))$.  While bubble shape modes break the spherical symmetry, and should therefore be fully characterised by the three-dimensional spherical modes $Y_k$, it is generally assumed that the two-dimensional projection of their shape remains axisymmetric [41–43], allowing a projection on $P_k$ along an axis of maximum symmetry $\theta_0(t)$. We choose this angle as the one for which $R(\theta_0+\theta,t)$ lies closest to $R(\theta_0-\theta,t)$. :}

\begin{equation}
       \zeta_k(t) = \frac{2 k + 1}{2} \int_{-1}^{1} \zeta(\theta,t) P_k (u) \, {\rm d}u \,,
\end{equation}
using $u = \cos(\theta - \theta_0)$. This projection actually defines two integration paths, one for each bubble hemisphere. We choose to fit each hemisphere separately and define $\zeta_k(t)$ as the average of the two. We finally compute the spectrograms $\hat{\zeta}_k(\nu,t)$ of the bubble shape modes:
\begin{equation}
    \hat{\zeta}_k(\nu,t) = \frac{2}{\Delta t} \left | \int_{t - \Delta t}^{t + \Delta t} \zeta_{k} (\tau) \exp (2 i \pi \nu \tau) \, {\rm d}\tau \right |
\end{equation}
We pay close attention to $\hat \zeta_k(t) = \hat \zeta_k(f/2,t)$, as $f/2$ is the frequency at which shape oscillations arise in Newtonian fluids \add{and Kelvin-Voigt materials}~\cite{Leighton1990,Murakami2020}. The time window size used to compute the spectrograms $\Delta t = 2 / f$ allows us to capture this component accurately.

\subsection{Characteristic quantities and dimensionless groups}
\label{sec:experiment_dimensionless}

The small bubbles ($75~\mu{\rm m} \leq R_0 \leq 300~\mu{\rm m}$) we consider in this article correspond to very small Bond-E\"otv\"os numbers,
\begin{equation}
    {\rm Bo} = \frac{\rho_{\rm l} g R_0^2}{\Gamma} \leq 0.01\,,
\end{equation}
and modest elasto-capillary numbers,
\begin{equation}
    {\rm El} = \frac{G R_0}{\Gamma} \leq 0.17\,.
\end{equation}
We then expect bubbles to remain spherical at rest, as hypothesised in Section~\ref{sec:theory}. The yielding parameter for such bubbles is \add{also} small,
\begin{equation}
    Y^{-1} = \frac{2 \rho_{\rm l} g R_0}{3 \sigma_{\rm Y}} = 0.26\,,
\end{equation}
since the critical value $Y_{\rm c}^{-1}$ needed to initiate the rise of spherical bubbles is $5.1$~\cite{Dimakopoulos2013}. We also provide a numerical estimate of the critical oscillation amplitude $\zeta_c$ required to initiate Carbopol yielding following Equation~\eqref{eq:yielding_vonMises}:
\begin{equation}
    \label{eq:yielding_vonMises_simplified}
    \zeta_c = \frac{1}{2 \sqrt{3}} \frac{\sigma_{\rm Y}}{G} = 0.043
\end{equation}
\add{We can finally compute the ratio of the elasto-plastic crossover time scale in the yielded material to that of the bubble oscillations, as defined in Section~\ref{sec:theory_yieldingcriterion}. We refer to it as the Deborah number of our experiments:
\begin{equation}
    \label{eq:deborahnumber_estimate}
    {\rm De} = 2 \pi f \left ( \frac{K}{G} \right )^{1/n} = 590 \gg 1.
\end{equation}
Hence, even if the material has yielded, it will remain predominantly elastic, and we do not expect the bubble oscillations dynamics to be affected by yielding. However, if the material has yielded due to bubble oscillations, irreversible bubble displacement may occur due to buoyancy and acoustic radiation forces, which are applied continuously during $N \geq 1000$ cycles, resulting in a time scale ratio $f/N (K/G)^{1/n} = {\rm De} / 2 \pi N$ below unity.}

We may lastly define the Péclet number comparing heat diffusion in the air to its advection due to bubble oscillations: ${\rm Pe} = 2 \pi f R_0^2/D = 160$. For this range of Péclet numbers, thermal dissipation is the dominant contribution to the damping term $\beta$ (see Appendix~\ref{sec:app:damping}) and the polytropic exponent is $\kappa = 1.30$~\cite{Prosperetti1977}. 


\section{Experimental Results}
\label{sec:results}

\subsection{Linear Response : General Observations}
\label{sec:results_linear}

\begin{figure*}[h]
    \centering
    \includegraphics[]{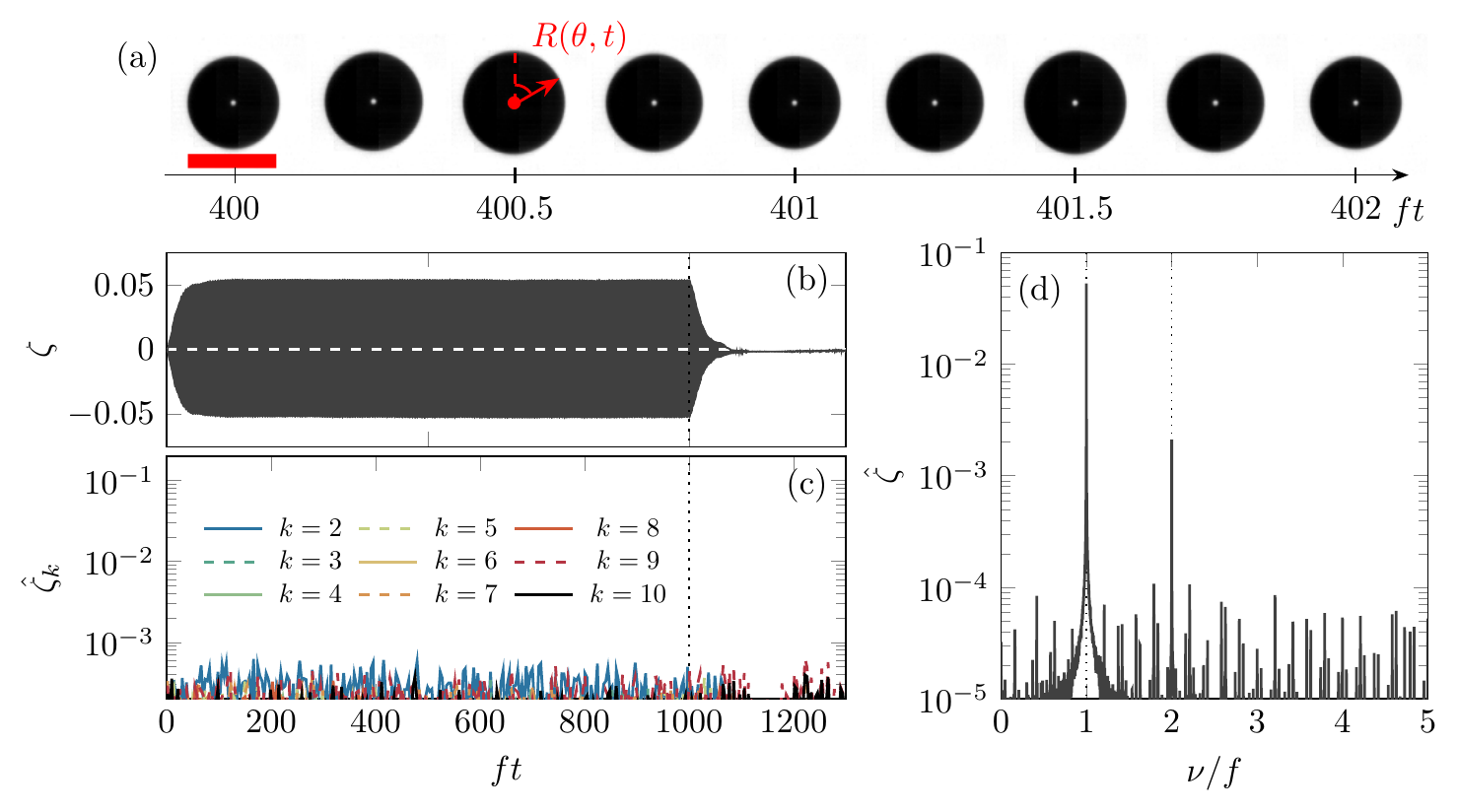}
    \caption{Linear oscillation of a bubble in a yield-stress fluid under acoustic excitation at a frequency $f = 21.5$~kHz. (a) Snapshots of the bubble oscillation during two oscillation cycles. The red scale bar is $250~\mu$m. (b) Time series of the relative departure from the initial radius $\zeta(t)$ as a function of time. Acoustic excitation starts at $ft = 0$ and stops at $ft = 1000$, marked by a dotted line. (c) Spectrogram $\hat \zeta_k(t)$ of the bubble shape modes for $k \leq 10$, showing no notable shape oscillation content at $f/2$. (d) Complex modulus of the Fourier transform of $\hat \zeta$.}
    \label{fig:analysis_linear}
\end{figure*}

We first show, for a typical experiment, the criteria we use to define spherical and linear bubble oscillations. Figure~\ref{fig:analysis_linear}(a) shows an image sequence of bubble oscillation after the end of the transient regime at location \textcircled{\footnotesize 1}. We also show in Figure~\ref{fig:analysis_linear}(b) the bubble oscillation amplitude $\zeta(t)$, which highlights the typical time scale $\sim 100/f$ needed for the transient state to vanish. The bubble radius at rest before and long after the oscillations are equal, ruling out any significant gas diffusion into or out of the bubble.
Figure~\ref{fig:analysis_linear}(c) confirms that bubbles remain spherical as the shape oscillation modes $\hat\zeta_k$ remain at a very low level before, during and after the acoustic excitation. We then set the noise threshold for shape oscillations to $10^{-3}$ for the rest of this article. Figure~\ref{fig:analysis_linear}(d) confirms the linearity in time of the bubble response. The bubble oscillation spectrum $\hat \zeta(\nu)$ shows a peak at $\nu = f$ and harmonic content is almost absent aside from a small peak at $\nu = 2f$. In all our experiments, spherical bubble oscillations are also linear in the time domain. 

\subsection{Resonance Curve}
\label{sec:results_resonancecurve}

\begin{figure}
    \centering
    \includegraphics{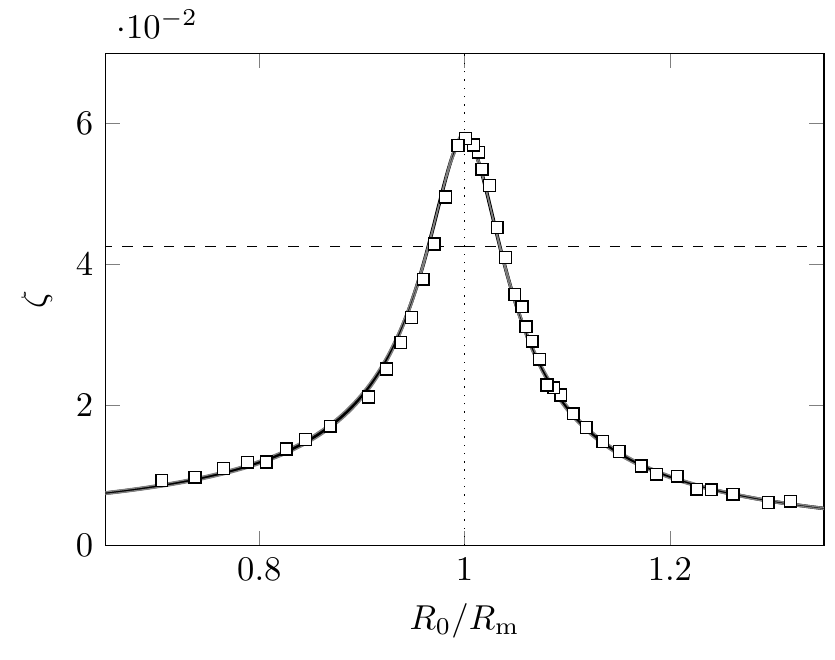}
    \caption{Resonance curve of a bubble in Carbopol obtained for an oscillation frequency of $21.5$~kHz and an acoustic pressure amplitude $p = 1.73$~kPa. The number of cycles has been set to 1000. The Minnaert resonance radius is $R_{\rm m} = 148~\mu$m. The dashed horizontal line represents the onset of Carbopol yielding deduced from Equation~\eqref{eq:yielding_vonMises}. Squares represent experimental data. The solid black line is a fit of the linear data following Equation~\eqref{eq:resonance_curve}, with two free parameters, the solvent viscosity $\eta_{\rm s}$ (included in the damping term $\beta$) and the applied pressure $p$. The 95\% confidence interval region is smaller than the size of the markers.}
    \label{fig:resonance_curve} 
\end{figure}

We then measure the resonance curve [Equation~\eqref{eq:resonance_curve}] of bubbles in the linear regime, sweeping over their initial radius $R_0$. Figure~\ref{fig:resonance_curve} highlights the excellent agreement between the measured oscillation amplitude $\zeta$ from a series of experiments conducted at a constant pressure amplitude and the prediction of Equation~\eqref{eq:resonance_curve}. We first verify that the fitted pressure amplitude $p = 1.73$~kPa matches independent pressure measurements using the hydrophone (data not shown). All experiments show neither any significant shape oscillation nor any non-linear behaviour in the time domain. \rem{The quality of the agreement allows us to fit the solvent viscosity $\eta_{\rm s} = 1.3 \pm 1.7$~mPa.s to the data, and we verify that this value is not overly sensitive to the details of the fitting procedure.} \add{By fitting the solvent viscosity to the data, and estimating uncertainties on thermal and acoustic damping from Equation~\eqref{eq:damping}, as detailed in ESI Section 1, we obtain an estimate for the solvent viscosity, $\eta_{\rm s} = 1.3 \pm 3.0$~mPa.s, compatible with the viscosity of water $\eta_0 = 1.0$~mPa.s, in agreement with the assumptions of the rheological model~\cite{Saramito2009}}

\rem{The interval includes the viscosity of water $\eta_0 = 1.0$~mPa.s used as the solvent here as expected from the  rheological model of Saramito \cite{Saramito2009} }. Other sets of data (not shown) performed at $19.45 \leq f \leq 29.2$~kHz are less precise but systematically include the viscosity of water in their confidence intervals. 

Close to $R_0 = R_{\rm m}$, 
the experiments in Figure~\ref{fig:resonance_curve} satisfy the yielding criterion $\zeta \geq \zeta_{\rm c}$, yet follow the exact same trend as the other experiments. \add{Material yielding has therefore no impact on the bubble dynamics. This result confirms the prediction made in Sections~\ref{sec:theory_yieldingcriterion} and~\ref{sec:experiment_dimensionless} that elastic stresses do not have time to relax in the yielded material and on the time scale of the oscillations}.

\rem{This result can be explained from the constitutive equations of our rheological model~\cite{Saramito2009,DeCorato2019} 
as the additional elasto-visco-plastic degree of freedom allowed above the yield stress is only activated when the deformation is applied for longer than a characteristic time $t_{\rm evp} = (K/G)^{1/n} \simeq 3.7 \times 10^{-3}$~s needed for the elastic stresses to relax. This time largely exceeds the typical oscillation timescale $1/f \simeq 5.0 \times 10^{-5}$~s. Hence, the bubble oscillations only probe the elastic part of the fluid response which does not change across the yield point. This result contrasts with classical rotational flow rheology (as seen in Appendix~\ref{sec:app:rheology}) in which shear is applied for a longer time and the viscous response is dominant.} \add{More surprisingly, we note that} \rem{We finally note that}  none of the experiments for which yielding is expected shows any \add{noticeable displacement of the centre of the bubble} \rem{irreversible bubble rise}. 

\subsection{Response to acoustic radiation forces}
\label{sec:results_motion}
\add{In Section~\ref{sec:results_resonancecurve} we did not observe any significant displacement of bubbles driven into oscillations at location \textcircled{\footnotesize 1}, where they are subject only to the buoyancy force. Next we test the effect of acoustic radiation forces [Equation~\eqref{eq:yielding_acousticradiation}] on bubble displacement, by looking at bubbles positioned at location \textcircled{\footnotesize 2} where they are subject also to acoustic pressure gradients.}
\rem{Since we did not observe any significant \rem{irreversible} motion of the bubble in the absence of acoustic pressure gradients (at location \textcircled{\footnotesize 1}) in Section~\ref{sec:results_resonancecurve}, we decide to drive bubbles into oscillations where such gradients are present (at location \textcircled{\footnotesize 2}) and initiate yielding through acoustic radiation forces [Equation~\eqref{eq:yielding_acousticradiation}].}

\begin{figure*}[t]
    \centering
    \includegraphics[scale=0.9]{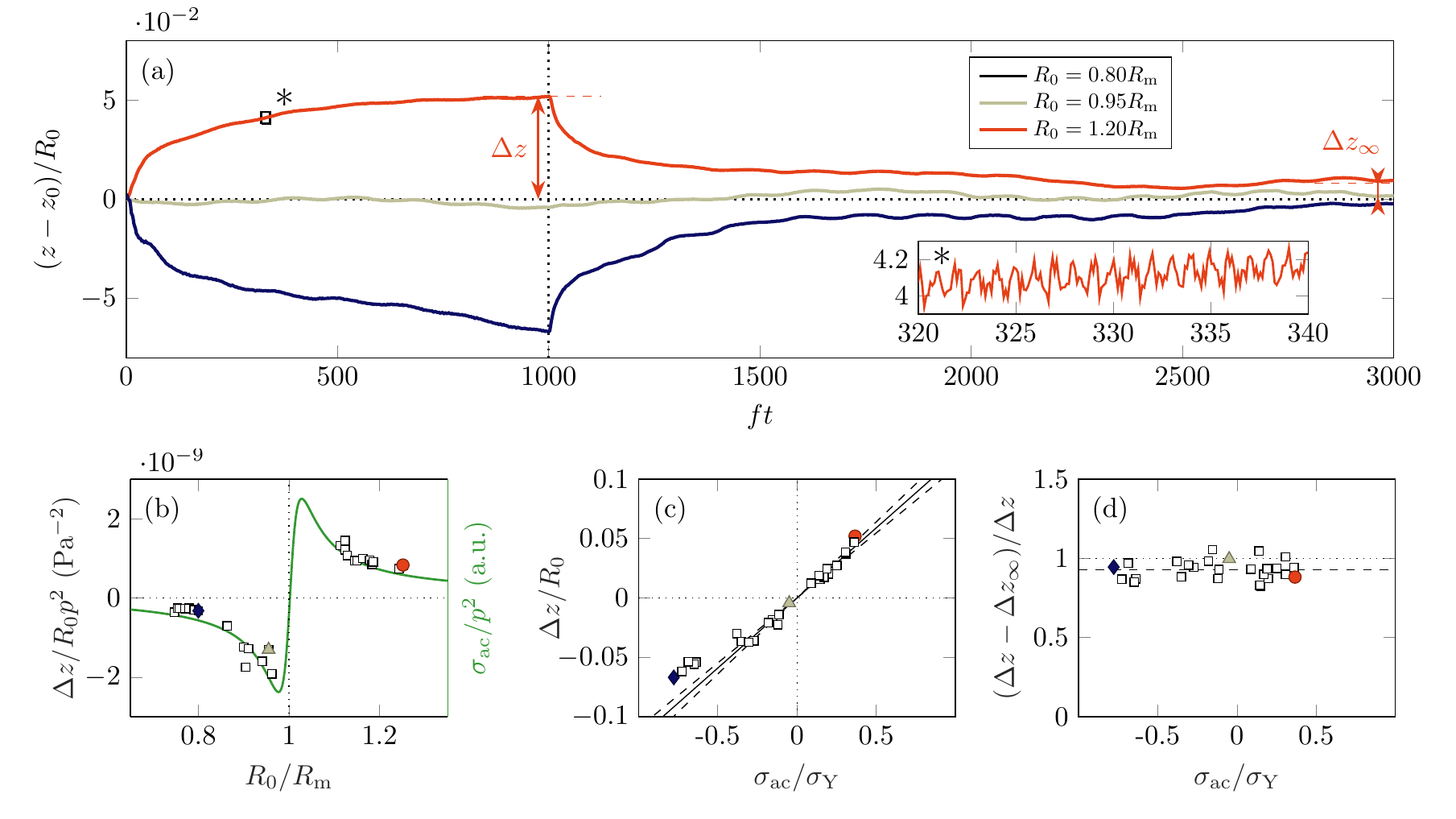}
    \caption{Displacement of the bubble centroid in the vertical direction for spherical bubble oscillation experiments conducted at $f = 19.45$~kHz. The Minnaert resonance radius is $R_{\rm m} = 163~\mu$m. 
    (a) Normalised bubble vertical displacement as a function of time for three experiments: $R_0 \simeq 1.2 R_{\rm m}$ (red), $R_0 \simeq 0.95 R_{\rm m}$ (light grey) and $R_0 \simeq 0.8 R_{\rm m}$ (blue). Data have been averaged over an oscillation period. Inset: raw temporal profile of the bubble vertical position highlighting the oscillating part of the signal. 
    (b) Typical strain applied by the bubble net motion for an applied pressure of $1$~Pa, $\Delta z/ R_0 p^2$ (markers), superposed with the expected acoustic radiation stress applied to the bubbles $\sigma_{\rm ac} / p^ 2$ (solid line, see main text) also for $p=1$~Pa. Note the separate $y$ axes.
    (c) Local rheology of the fluid: typical measured strain $\Delta z/R_0$ plotted as a function of the acoustic radiation stress normalised by the yield stress $\sigma_{\rm ac} / \sigma_{\rm Y}$. The solid line is the best linear fit of the data for $ |\sigma_{\rm ac} / \sigma_{\rm Y} | < 0.5$ and the dashed lines show the linear fits from the boundaries of the 95\% confidence interval. 
    (d) Recovery after strain of the bubble: the amount of recovered strain $(\Delta z - \Delta z_{\infty}) / \Delta z$ long after the end of acoustic excitation is plotted as a function of the normalised acoustic radiation stress. In Panels (b), (c) and (d), the red, light grey and red markers correspond to the three lines of Panel (a).}
    \label{fig:movement_z}
\end{figure*}

Figure~\ref{fig:movement_z}(a) shows the vertical position of the bubble centroid $z(t)$ for three experiments. Bubbles smaller (respectively larger) than the resonant radius $R_{\rm m}$ 
show a net downwards (respectively upwards) motion towards the pressure anti-node (respectively pressure node), in line with the change of sign of $\cos(\phi)$ in Equation~\eqref{eq:phase_lag}. The inset of Figure~\ref{fig:movement_z}(a) highlights the zero-average oscillatory part of the acoustic radiation forces [averaged out in Equation~\eqref{eq:bjerknes}], clearly noticeable and superposed with the slower displacement related to the Bjerknes force.

Bubble trajectories are non-trivial: they cannot be fitted by a simple exponential law related to the \add{Kelvin-Voigt} solid visco-elastic relaxation time $f t_{\rm KV} = \eta_{\rm s}/ G$, \add{which amounts to less than one oscillation cycle, nor to the time needed for the transient regime to die out, which corresponds to around $100$ cycles, or even the typical elasto-plastic relaxation time of the yielded material, given by ${\rm De} / 2 \pi N = 1$, also close to $N=100$ cycles.}
\rem{the elasto-visco-plastic time $t_{\rm evp} \sim 100 / f$ or the time needed for the transient regime to die out, also of the order $100 / f$} (see Figure~\ref{fig:analysis_linear}). We can rule out viscous or plastic responses of the fluid as the bubble centroid does not reach a constant, finite velocity ${\rm d}z/{\rm d}t$. They are however not long enough to be completely conclusive regarding more complex, non-linear responses of the fluid, such as creep~\cite{Lidon2017a}.

Figure~\ref{fig:movement_z}(b) examines the sensitivity of bubbles to acoustic radiation forces as a function of their size, defined as the normalised displacement $\Delta z / R_0 p^2$ as Bjerknes forces are quadratic in pressure amplitude (see Section~\ref{sec:theory_acousticradiation}). Our experimental data superposes well with the theoretical expression for the average stress applied onto the bubble $\sigma_{\rm ac} / p^2$ from Equations~\eqref{eq:resonance_curve}, \eqref{eq:phase_lag} and \eqref{eq:bjerknes}, which suggests a linear relation between bubble stress and strain. 

Figure~\ref{fig:movement_z}(c) directly plots the acoustic radiation strain $\Delta z/R_0$ as a function of the corresponding stress, normalised here by the yield stress $\sigma_{\rm Y}$. We compute here the stress using experimental values of $\zeta$, $R_0$, $p$ and $|{\bm \nabla} \! p|$ and we choose $\cos(\phi)$ based on Equation~\eqref{eq:phase_lag}. The data confirms the linear trend suggested from Figure~\ref{fig:movement_z}(b) at low applied stresses and shows a noticeable non-linear deviation for higher stresses. As yielding due to the oscillation amplitude has no impact on the bubble mobility (see ESI Section 3), this deviation may only stem from a non-linear behaviour of the emission setup or non-linear elasticity of the Carbopol. We measure the slope of the linear trend at low stress in Figure~\ref{fig:movement_z}(c) to extract an estimate of the linear elastic modulus of the surrounding medium  $G = 44.4 \pm 3.5$~Pa, following Equation~\eqref{eq:displacement_linelastic}. This value is in fair agreement with that obtained from bulk oscillatory rheology, $G = 36.0$~Pa. 

Figure~\ref{fig:movement_z}(d) shows the recovered strain $2000$ cycles after the end of the acoustic excitation. The recovery is close to $100\%$ for all experiments, which confirms the elastic nature of the deformation shown in Figure~\ref{fig:movement_z}(c) expected for experiments conducted for 
$\sigma_{\rm ac} / \sigma_{\rm Y} \leq 5.1$. Irreversible bubble motion can then only be achieved for higher applied pressure and oscillation amplitude $\zeta$. As we will see in Section~\ref{sec:results_shapeinstab}, we could not perform such experiments due to the onset of bubble shape oscillations.

\subsection{Shape oscillations}
\label{sec:results_shapeinstab}

\subsubsection{\add{Critical pressure and} observed modes}

\rem{In Newtonian fluids and soft solids, shape oscillations of mode number $k$ may grow when the applied pressure $p$ (or the linear oscillation amplitude $\zeta$) exceeds a critical value $p_{\rm c}^k$ (or $\zeta_{\rm c}^k$)~\cite{Maksimov2001,Murakami2020}. This critical pressure itself depends on the bubble radius and usually presents two local minima: the first is independent of $k$ and is located at the Minnaert resonance of the bubble, while the second is the natural oscillation frequency of the shape mode $k$, given by~\cite{Lamb1932,Murakami2020}}

\rem{Recent experiments in Newtonian fluids~\cite{Versluis2010,Cleve2019} are in reasonable agreement with the theory, especially considering the numerous hypotheses used to derive the critical pressure $p_{\rm c}^k$. Very few experiments~\cite{Hamaguchi2015} or even numerical simulations~\cite{Foteinopoulou2010} are available to validate the prediction of Equation~\eqref{eq:lambshapemodes} for viscoelastic materials. In addition, no theoretical framework has been specifically derived to predict shape oscillations in yield-stress fluids. In the rest of this Section, we will therefore compare our shape oscillation observations with the theory developed for soft viscoelastic solids~\cite{Murakami2020}.}

\add{In Newtonian fluids and soft solids, shape oscillations of mode number $k$ may grow when the applied pressure $p$  exceeds a critical value $p_{{\rm c},k}$, which depends on $R_0$, the applied frequency $f$ and the material properties~\cite{Maksimov2001,Murakami2020}. For a fixed driving frequency $f$, these predictions define regions in the $(R_0, p)$ plane in which bubble oscillations either remain spherical, allow the growth of a single shape mode $k$, or allow multiple shape modes. Linear instability predictions for shape oscillations in Kelvin-Voigt soft solids have recently been derived~\cite{Murakami2020}; they are recalled in Appendix~\ref{sec:app:critshapeamp}. In Newtonian fluids, experimental results match the linear instability predictions fairly well~\cite{Versluis2010,Mekki-Berrada2016,Cleve2019}. In contrast, numerical~\cite{Foteinopoulou2010} and experimental~\cite{Hamaguchi2015} data on the onset of bubble shape oscillations in non-Newtonian fluids are  scarce and not yet conclusive.}

\begin{figure*}
    \centering
    \includegraphics{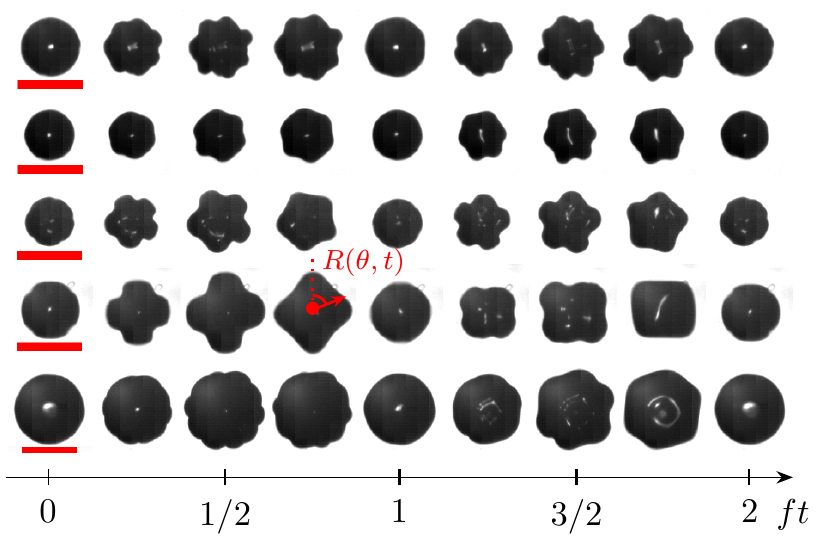}
    \caption{Shape oscillation modes observed in our experiments, shown for two oscillation periods ($2/f$). From top to bottom, the applied frequency is $f = 27.94$~kHz, $f = 29.2$~kHz, $f = 27.94$~kHz, $f = 19.45$~kHz and $f = 19.45$~kHz. The first four rows show shape oscillations with clear modes $k = 7$, $6$, $5$ and $4$ respectively. The last row shows an experiment for which shape oscillations are clearly detected but the corresponding mode $k$ is difficult to ascertain. The red scale bars are $300~\mu$m.}
    \label{fig:shape_oscillations}
\end{figure*}

\rem{Figure~5(a) highlights four shape oscillation modes $4 \leq k \leq 7$ that have been clearly identified in our experiments at location \textcircled{\footnotesize 1}. Most experimental data are sufficiently clear to define unambiguously a shape mode number $k$, with several exceptions, as shown in the last row of Figure~5(a). In all cases, the frequency of the shape oscillations is $f/2$, confirming that they result from a sub-harmonic instability also in yield-stress fluids.}

\rem{Figure~5(b) shows that, at a fixed applied frequency $f$, the shape mode number $k$ increases with $R_0$ both in experiments and in Equation~\eqref{eq:lambshapemodes}. The experimental data presents a large dispersion, especially for the $k = 6$ mode. In addition, the shape modes $k$ predicted by Equation~\eqref{eq:lambshapemodes} grow significantly slower with $R_0$ than the overall experimental trend. Including the elastic modulus $G = 44.4$~Pa measured in Section~\ref{sec:results_motion} has a very limited influence on this prediction given the small elasto-capillary numbers ${\rm El}$ of our experiments.}

\add{Figure~\ref{fig:shape_oscillations} highlights four shape oscillation modes $4 \leq k \leq 7$ that have been clearly identified in experiments at location \textcircled{\footnotesize 1}. Less than half of the experimental data is sufficiently clear to define unambiguously a shape mode number $k$. Several experiments (see last row of Figure~\ref{fig:shape_oscillations}) instead show a complex outline, which likely results from the projection in the imaging plane of a three-dimensional mode $Y_k^m$ with $m \neq \{0,k,-k\}$ and a random orientation. In all cases, the frequency of the shape oscillations is $f/2$, confirming that shape oscillations also result from a sub-harmonic instability in yield-stress fluids.}

\begin{figure*}
    \centering
    \includegraphics{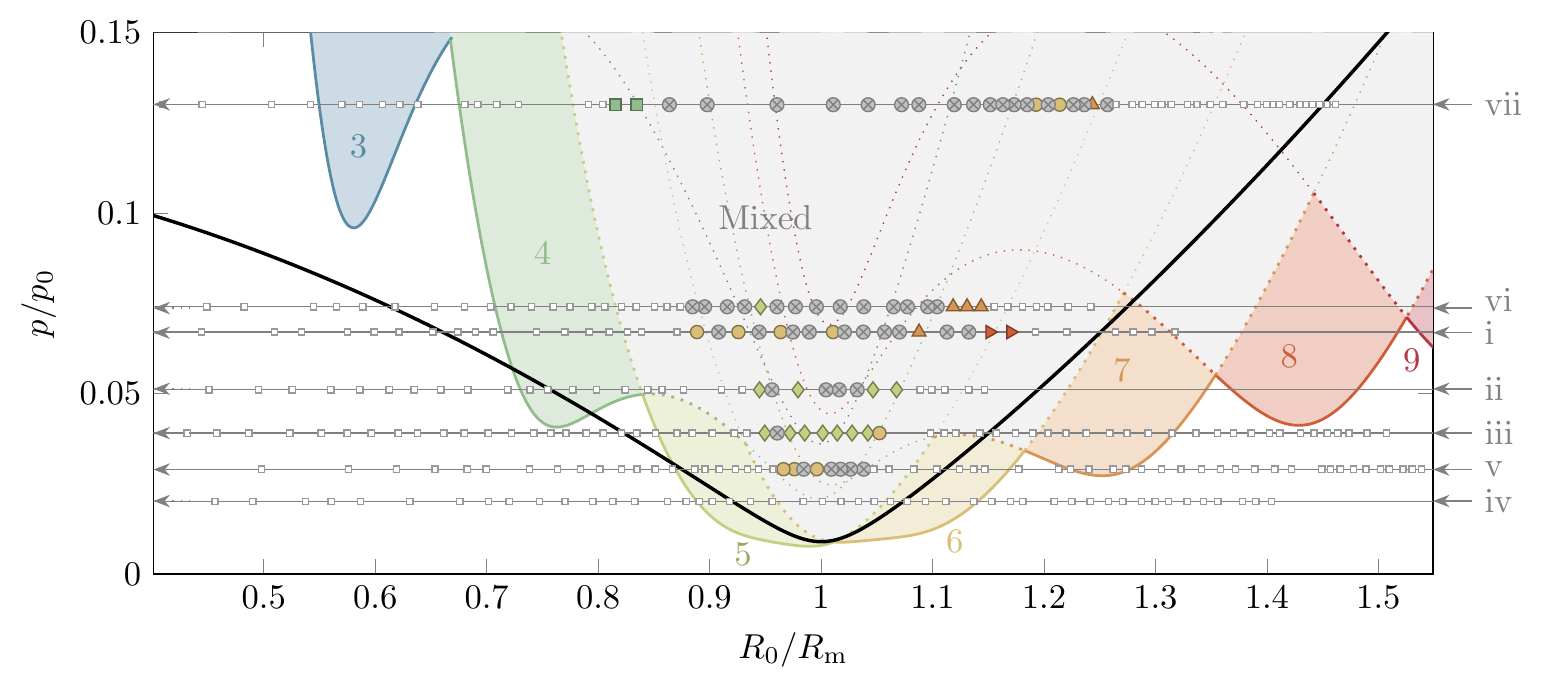}
    \caption{Phase diagram for bubble shape oscillations as a function of the applied pressure $p/p_0$ and the initial radius of bubbles $R_0/R_{\rm m}$ for an excitation frequency $f = 22.5$~kHz. \add{Experimental data have been acquired throughout the dissolution of seven bubbles, identified with roman numerals i to vii from earliest to latest data series. As the applied pressure $p$ is kept constant for each bubble, the seven data series form horizontal series of points, starting from high values of $R_0$ and ending for low values of $R_0$, following the direction of the grey arrows.
    Individual acquisitions for each data series} are shown as symbols, \rem{with the same marker and colour conventions as in Figure~\ref{fig:shape_oscillations}(b)} \add{The modes that we could define without any ambiguity are plotted as red rightwards pointing triangles ($k=8$), orange upwards pointing triangles ($k=7$), yellow circles ($k = 6$), light green diamonds ($k=5$) and green squares ($k=4$). Most acquisitions with shape oscillations have a non-clear mode [e.g. last row of Figure~\ref{fig:shape_oscillations}]}; they are plotted as crossed circles. The small white squares represent stable spherical oscillations. The critical pressure for each shape mode, computed from Ref.~\citenum{Murakami2020} \add{and Equation~\eqref{eq:resonance_curve}}, is shown as a line with the same colour coding as the experiments. Above these lines lie coloured regions where only one oscillation mode $k$ can grow, and a broader grey region where multiple shape modes may grow. The black solid line depicts the threshold for fluid yielding defined combining Equations~\eqref{eq:resonance_curve} and \eqref{eq:yielding_vonMises}.}
    \label{fig:pressure_instability}
\end{figure*}

\rem{We report in Figure~\ref{fig:pressure_instability} the bubble shape modes observed for seven bubbles driven at different but fixed pressure amplitude $p$ throughout their slow dissolution. Each dissolving bubble --which corresponds to a horizontal line in Figure~\ref{fig:pressure_instability}-- confirms that a larger $R_0$ promotes higher shape modes. We observe a significant variability between different bubbles driven at similar pressures (e.g., the two lines at $p/p_0 \simeq 0.065$ and $0.075$). More specifically, shape modes $k$ are consistently higher throughout the life time of a bubble when the first shape oscillation mode it experiences -- the rightmost colour symbol on each line -- is also high.} 

We report in Figure~\ref{fig:pressure_instability} the shape oscillations observed as a function of both $R_0$ and $p$ for seven \add{slowly dissolving bubbles, identified by a roman numeral from i to vii. Multiple acquisitions have been conducted on each bubble, with the pressure $p$ kept constant throughout their dissolution. The critical pressure of shape oscillations reaches a single local minimum close to $R_0 = R_{\rm m}$; further away from $R_{\rm m}$, it quickly grows and ultimately exceeds the maximum pressure achieved in our setup for $R_0 \leq 0.8 R_{\rm m}$ and $R_0 \geq 1.3 R_{\rm m}$. Our data indicates that the shape number $k$ in our experiments increases with $R_0$, in qualitative agreement with models~\cite{Maksimov2001,Murakami2020} and experiments in Newtonian fluids~\cite{Versluis2010,Cleve2019}.}

We overlay in Figure~\ref{fig:pressure_instability} the predicted critical pressure \add{$p_{{\rm c}, k}$} derived \add{in Appendix~\ref{sec:app:critshapeamp}} by combining Equation~\eqref{eq:resonance_curve} and the critical bubble oscillation amplitude \add{$\zeta_{{\rm c}, k}$} above which spherical oscillations are linearly unstable. We choose the value of the viscosity we fitted in Section~\ref{sec:results_resonancecurve}, $\eta_{\rm s} = 1.3$~mPa.s and the elastic modulus measured in Section~\ref{sec:results_motion}, $G = 44.4$~Pa. A large amount of experiments shows stable spherical oscillations whereas the model predicts they are linearly unstable with respect to shape oscillation modes $4$ to $8$. \add{The model however correctly predicts that modes $5$ and $6$ are favoured for $R \simeq R_{\rm m}$ in agreement with the low values of $p_{{\rm c},5}$ and $p_{{\rm c},6}$ in this region}. \rem{We show in Figure~SI.3 that the elastic term has a very limited impact on the critical pressure $p_{\rm c}^k$ and that artificially increasing the viscosity up to $\eta_{\rm s}=6.0$~mPa.s improves the agreement between the prediction and the experiments. However, this viscosity falls out of the viscosity interval obtained by fitting the resonance curve and the optimal agreement remains unsatisfactory.}

\subsubsection{Impact on net bubble motion}
In Newtonian fluids, bubble shape oscillations are closely related to an unpredictable, dancing~\cite{Doinikov2004} motion of their centre of gravity. This motion stems from a non-linear interaction with both spherical oscillations and other shape modes~\cite{Doinikov2004}. In the context of bubble removal, we wish to understand the impact of shape oscillations and dancing motion on the ability of ultrasound devices to push and pull bubbles irreversibly in yield-stress fluids.

\begin{figure*}
    \centering
    \includegraphics[trim=12pt 0pt 12pt 0pt, clip, scale=0.90]{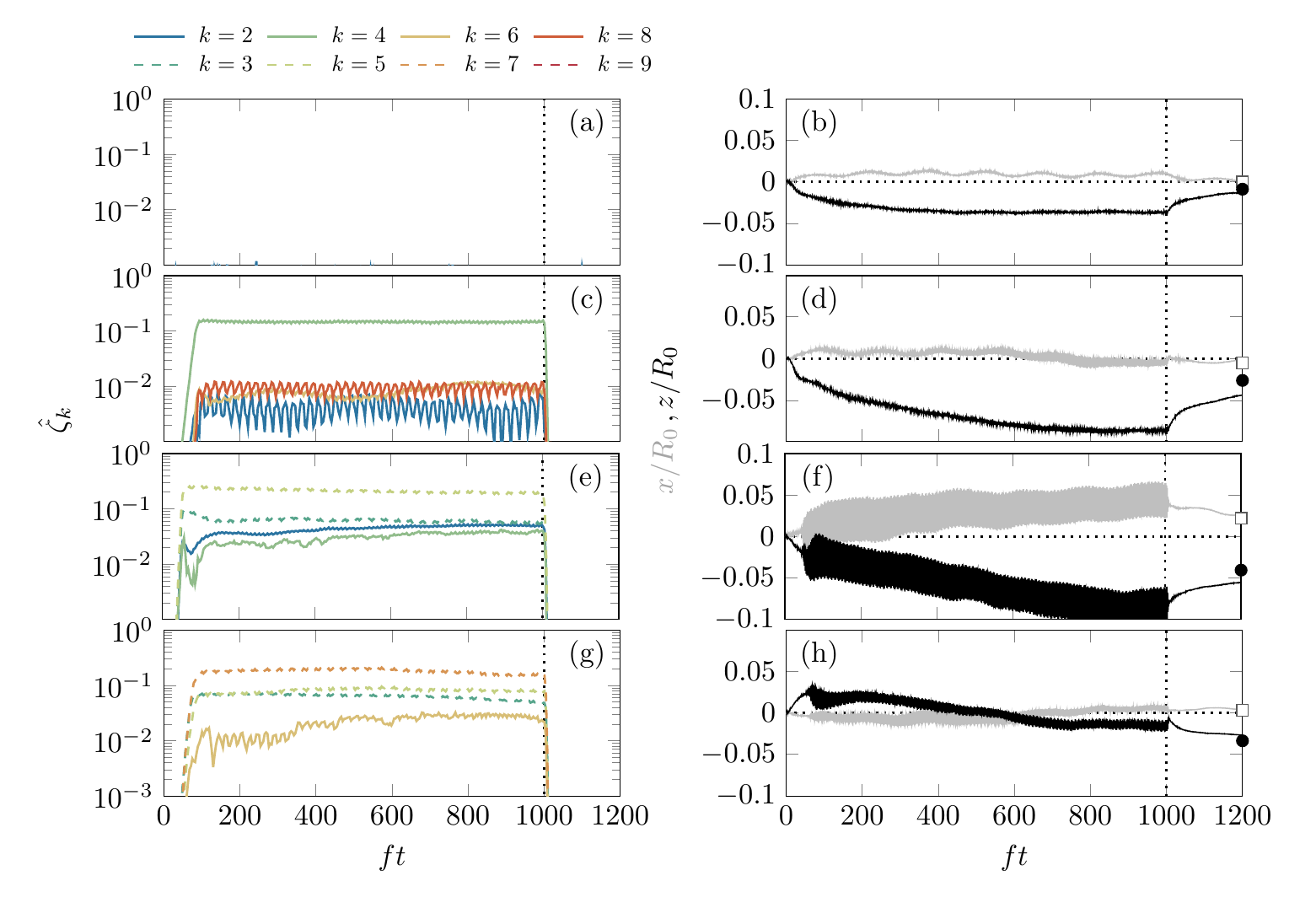}
    \caption{Analysing experiments for which one or several shape oscillation modes are observed and steady. Panels (a)-(c)-(e)-(g), spectrograms of the recorded shape mode intensity as a function of time for an experiment with no shape oscillation (a), a \add{clearly identified mode $k = 4$ (c) and two experiments (e,g) for which the shape modes are not clearly identified.}    
    \rem{strongly dominant mode 4 (c), a dominant mode 5 (e) and a weakly dominant mode 7 (g).} The acoustic excitation starts at $t = 0$ and stops at $f t = 1000$, marked by a dotted line. Only the four most prominent projection modes are shown. Panels (b)-(d)-(f)-(h), net bubble motion corresponding to the spectrograms of Panels (a)-(c)-(e)-(g). The grey line shows the relative displacement along the $x$ (horizontal) axis while the black line shows its $z$ (vertical) counterpart. The white square and the black circle respectively represent the position of the bubble in the $x$ and $z$ axes at $f t \simeq 3100$, long after the oscillations have stopped.}
    \label{fig:bubble_trajectories}
\end{figure*}

Figure~\ref{fig:bubble_trajectories} shows the strong impact of shape oscillations on bubble motion. The 
first three bubbles [Figure~\ref{fig:bubble_trajectories} (a-f)] show motion towards an antinode in agreement with their initial size $R_0 \leq R_{\rm m}$. 
\add{The presence of a clear shape mode enhances bubble mobility, as shown in Figure~\ref{fig:bubble_trajectories}(c-d) for $k=4$. We also observe spurious motion in the direction transverse to the pressure gradient when a single bubble shape mode $k$ is no longer clearly identified [as seen in Figure~\ref{fig:bubble_trajectories}(e-f)]. 
We also have observed reversals of the bubble direction of motion following the onset of shape oscillations. [Figure~\ref{fig:bubble_trajectories}(g-h)]. In general we conclude that while shape oscillations increase bubble displacement, the direction of  motion can no longer be controlled.}

\rem{However, this motion is enhanced when a single shape mode is present, or accompanied by an unexpected displacement in a second direction [Figure~\ref{fig:bubble_trajectories}(e-f)] in the case of multiple shape modes. Additionally, multiple shape modes can reverse the direction of motion, as seen here for a bubble initially larger than the resonant size. In this case, the bubble relaxation drives the bubble further away from its initial position. In the general case, the bubble shape dynamics is even more complex than what we report in Figure~\ref{fig:bubble_trajectories} and the associated bubble displacement cannot be predicted even qualitatively.}

\section{Discussion}
\label{sec:discussion}

\subsubsection*{Precision and relevance of the solvent viscosity measurement}
The experimental bubble oscillation amplitude $\zeta$ in the linear regime may be fitted to the theoretical resonance curve to extract the oscillation damping parameter $\beta$. After carefully subtracting from $\beta$ the dominant acoustic and thermal contributions, we measure a fluid viscosity $\eta_{\rm s} = 1.3$~mPa.s. \add{An analysis of the fitting procedure and the uncertainties on $\beta$ shows that this value is not statistically different from the viscosity of water} \rem{in surprising agreement with that of water} used as the solvent here, and in agreement with the rheological model~\cite{Saramito2009,DeCorato2019}. This low value \add{seems} surprising considering the bulk oscillatory rheology of Carbopol (see Appendix~\ref{sec:app:rheology}) which rather suggests a viscosity $G''/2 \pi f \simeq 1$~Pa.s in the linear regime for low frequencies $f=1$~Hz, while the \add{Kelvin-Voigt} model we use assumes a constant viscosity below yielding for all frequencies.

\add{Indeed, real hydrogels and yield-stress fluids under oscillatory shear do not show a constant viscosity as a function of $f$: classical rheological measurements show that their loss modulus $G''$ behaves as a constant or as slowly increasing power laws of $f$~\cite{Jaishankar2013} leading to a decreasing viscosity $G''/ 2 \pi f$. These power law scalings may be reproduced by fractional derivative models~\cite{Jaishankar2013} but their microscopic origin remain insufficiently understood~\cite{Nicolas2018}. The value of the viscosity deduced from $G''$ in oscillatory rheology may therefore not be particularly meaningful. In contrast, viscous or close-to-viscous scaling of the stress has been experimentally observed in yield-stress fluids at high frequencies and strain rates~\cite{Mason2000,Caggioni2020}. At such frequencies, dissipation due to the solvent, scaling as $\eta_{\rm s} f$, may become the dominant contribution to $G''$, and the material could then recover a Kelvin-Voigt rheology. Our results suggest that bubble oscillations experiments fit into this high-frequency limit and allow a proper measurement of the solvent viscosity. }

\rem{Very recent rheological models~\cite{Fraggedakis2016,Dimitriou2019} now allow plastic deformation in transient flows. In line with our observations, they predict a relatively high loss modulus at the bulk rheology frequency and a sharp decrease in the viscosity at higher frequencies i.e. below the characteristic time scale of the dynamics of the yield stress.} 

\subsubsection*{Linear response to Bjerknes forces}
We have used in Section~\ref{sec:results_motion} the constant (or zero-frequency) part of the acoustic radiation force to perform an equivalent of step-stress tests, but at a local scale $R_0$ \rem{and on a relatively short timescale $N/f \simeq 0.05$~s}. For moderate acoustic stresses $\sigma_{\rm ac} \leq \sigma_{\rm Y}$, we measure a linear strain-stress relation at the end of oscillations, from which we deduce an independent measurement of the local linear elastic modulus of the fluid below yielding, $G = 44.4 \pm 3.5$~Pa, comparable to that obtained using bulk rheology, $G = 36$~Pa. All quantities used to derive $G$ are either directly measured or estimated from the resonance curve: hence, in contrast with previous works~\cite{Lidon2019}, our measurement is truly independent from bulk rheology.
The complex time dependence of the displacement shown in Figure~\ref{fig:movement_z}(a) is reminiscent of creep behaviour~\cite{Lidon2017a}. Creep is however usually associated to irreversible strain and a non-linear stress-strain relation in bulk rheology experiments, both of which are not observed here. Interestingly, fully reversible creep motion up to the yield point has also been reported in experiments in which acoustic radiation forces are used to push small spheres~\cite{Lidon2019}. The relatively small pressure gradients applied in our experiments according to Equation~\eqref{eq:yielding_acousticradiation} then cannot alone initiate bubble rise.  \add{Performing experiments of longer duration may reveal whether the response to acoustic radiation forces indeed follows a power law or an exponential profile with time, which could be helpful to validate the recent, advanced models of yield-stress fluids~\cite{Dimitriou2019,Nicolas2018}}.

\subsubsection*{Absence of irreversible rising motion}

Several experiments satisfy the bubble oscillation yielding criterion $\zeta \geq \zeta_{\rm c}$ and apply acoustic radiation stresses $\sigma_{\rm ac}$ comparable with the yield stress \add{for a sufficiently long time to let elastic stresses relax.} Yet, they do not suffice to induce irreversible bubble motion and we do not observe the finite average rising speed predicted in the recent numerical simulations of Ref.~\citenum{Karapetsas2019}. The yielding criterion $\zeta_{\rm c} = 0.043$ we have derived is then a necessary condition, but not sufficient, to induce bubble rise at a useful rate for removal applications. 

\rem{Several factors can explain the absence of bubble motion despite $\zeta \geq \zeta_{\rm c}$. First, the size of the yielded region around the bubble remains small for all our experiments for which $\zeta \leq 0.08$ and rising motion is therefore strongly confined~\cite{DeCorato2019}}
\add{One explanation for this lack of irreversible motion is that the steady-state bubble rise velocity is too small to be observed. Firstly, the yielded region remains under $1.25$ times the size of the bubble radius, increasing drag by a factor $40$ compared to the unconfined case~\cite{Happel1983}. Secondly, the plastic viscosity in the yielded material stays significantly higher than the solvent viscosity. 
The corresponding rising velocities may therefore be too small to be resolved in experiment.}

\add{Additional factors may prevent irreversible rising motion. For instance, the von Mises yield criterion [Equation~\eqref{eq:yielding_vonMises}] has been shown to fail} \rem{It could also be related to a failure of the von Mises yield criterion [Equation~\eqref{eq:yielding_vonMises}]}
for bulk yielding in extension, as already reported in other simple yield-stress fluids~\citep{Niedzwiedz2010,Zhang2018,Varchanis2020}. Another possibility lies in finite-size effects given the relatively small size of the bubble compared to the constitutive elements of Carbopol. Local restructuration around slowly-growing bubbles has been recently evidenced in sparse networks of microfibrillated cellulose, which impacts their bubble retention capacity~\citep{Song2019a,*Song2019b}. It is difficult to know at the moment whether this scenario applies in our case, since Carbopol is soft-jammed and isotropic and the strain rates at play are high. We may finally question the relevance of the very notions of yielding and unyielding in our experiment since the oscillation timescale $1/f$ can be below that of \add{the 
microscopic plastic rearrangements} \rem{plastic relaxation events} used in yield-stress fluids models~\cite{Dimitriou2019,Nicolas2018}.

\subsection*{Nature and onset of shape oscillations}

\add{The critical pressure $p_{\rm c}^k$ above which we experimentally observe shape oscillations is significantly higher in Carbopol thant what we expect from a linear instability analysis in Kelvin-Voigt materials~\cite{Murakami2020} if we use the fluid properties we derived in Sections~\ref{sec:results_resonancecurve} and \ref{sec:results_motion}.} 
\add{Yield-stress fluids are known to exhibit residual stresses at rest, with unknown spatial distribution. We expect non-homogeneous residual stresses around the bubble to impact bubble shape oscillations by altering the critical pressure $p_{{\rm c},k}$ depending on the compatibility between the geometry of the shape modes and that of the residual stresses. Further analysis of the bubble shapes, conducted in ESI Section 4, shows that residual stresses induce a very small ($0.5\%$) residual deformation of the bubble at rest. Under acoustic excitation, the bubble shape modes do neither respect the orientation nor the symmetry of these residual deformations. Hence we do not observe any direct impact of residual stresses on bubble shape oscillations even though we cannot rule out their influence. 
Using a solvent with a higher viscosity in experiments would be particularly helpful to either reconcile experimental data with the linear instability model~\cite{Murakami2020} or to prove that it is not applicable to yield-stress fluids.}

\rem{The shape oscillations we report in the experiments at high pressure amplitudes differ substantially in shape number $k$ and critical pressure $p_{k, c}$ from the available predictions derived for soft elastic solids. We believe residual stresses explain the variability in shape modes observed in Figure~\ref{fig:shape_oscillations}(b) and their persistence for a given bubble in Figure~\ref{fig:pressure_instability}. The origin of the general departure of our data from Equation~\eqref{eq:lambshapemodes} is less clear, especially given the relatively low elastocapillary numbers $GR_0/\Gamma$ at play in our experiments. The critical pressure above which shape oscillations are observed is higher in the experiments than in the model~\cite{Murakami2020} if we use the solvent viscosity $\eta_{\rm s}$ and the elastic modulus derived in Sections~\ref{sec:results_resonancecurve} and~\ref{sec:results_motion}. Increasing artificially the viscosity of the solvent in the model improves the agreement with the experimental data but this agreement is never satisfactory. The lack of good agreement with the model~\cite{Murakami2020} is quite surprising given their correct prediction of shape modes for another soft solid --gelatin--. We speculate that dissipative processes are more complex in yield-stress fluids than in Newtonian fluids and ideal viscoelastic solids. For instance, the spatial cooperativity of dissipative processes in yield-stress fluids~\cite{Goyon2008} may have to be included to make accurate predictions for $p_{\rm c}^k$.}


\subsubsection*{Consequences on acoustic bubble removal performance}
Shape oscillations imply unpredictable bubble motion that inevitably reduces the efficiency of any directed motion induced by acoustic radiation forces or bubble buoyancy. We notice that the window of operation for bubble removal, lying above the black line and below the coloured symbols of Figure~\ref{fig:pressure_instability}, is limited especially since the yielding criterion of Equation~\eqref{eq:yielding_vonMises} does not warrant bubble rise. Bubble removal using acoustic excitation in Carbopol \rem{should} \add{could} then be performed using stronger pressure gradients, for instance using focused ultrasound beams. We suspect Carbopol is particularly resistant to the removal process due to its very wide linear elastic regime, as shown in bulk rheology (Appendix~\ref{sec:app:rheology}). Bubble removal should be easier in almost any other yield-stress fluid as they break down under much smaller strains \citep{Andrade2019,Saha2020}. Increasing $\eta_{\rm s}$ could also improve bubble removal by raising the critical pressure at which shape oscillations arise. 

\section{Conclusion}
\label{sec:conclusion}
In this article, we have investigated how a small bubble oscillating at a high frequency interacts with Carbopol, a model yield-stress fluid. Bubbles of different sizes allow us to perform bubble spectroscopy~\cite{vanderMeer2007,Hamaguchi2015} and extract a viscosity $\eta_{\rm s} = 1.3$~mPa.s of the fluid at high frequency and for a finite extensional deformation, in agreement with the solvent viscosity of water and as expected from a previous numerical study~\cite{DeCorato2019}. We have also used pressure gradients to apply acoustic radiation forces on bubbles, from which we measure the local linear shear modulus of the fluid $G = 44.4$~Pa, in fair agreement with bulk rheology. As long as the oscillations remain spherical, bubble motion is fully reversible given the range of acoustic radiation stresses $|\sigma_{\rm ac}| \leq \sigma_{\rm Y}$ achieved in our experiment. In particular, motion reversibility \add{appears} unaffected by the oscillatory yielding criterion derived by \citet{DeCorato2019}.

Experiments performed at higher pressure always resulted in non-spherical shape oscillations. \rem{The non-Newtonian nature of the fluid --notably, the presence of residual stresses-- alters the shape mode selection known for Newtonian fluids and the critical pressure at which these modes are observed} 
As shape oscillations result in an unpredictable bubble motion in all directions, acoustic bubble removal is quite inefficient in Carbopol. \add{Future studies should explore the applicability of acoustic bubble removal in more fragile networks, corresponding to a wide range of attractive colloidal and athermal yield stress fluids in which spherical bubble oscillations largely beyond the yield point are possible, resulting in a strong decrease of both bubble confinement and plastic viscosity during its assisted motion.} 
\rem{Nevertheless, we believe that the technique can still be used on more fragile networks, i.e. for a broad range of attractive colloidal and athermal yield-stress fluids.}

\section*{Conflicts of Interest}
There are no conflicts to declare.

\section*{Acknowledgements}
The authors wish to thank M. De Corato, J. Tsamopoulos and Y. Dimakopoulos for stimulating discussions and their critical reading of the paper. They also thank D. Baresch for his help with the design of the experimental setup. This work is supported by European Research Council Starting Grant No. 639221 (V.G.).

\clearpage

\appendix

\section*{Appendix}

\section{Carbopol Rheology}
\label{sec:app:rheology}

\begin{figure}
    \centering
    \includegraphics{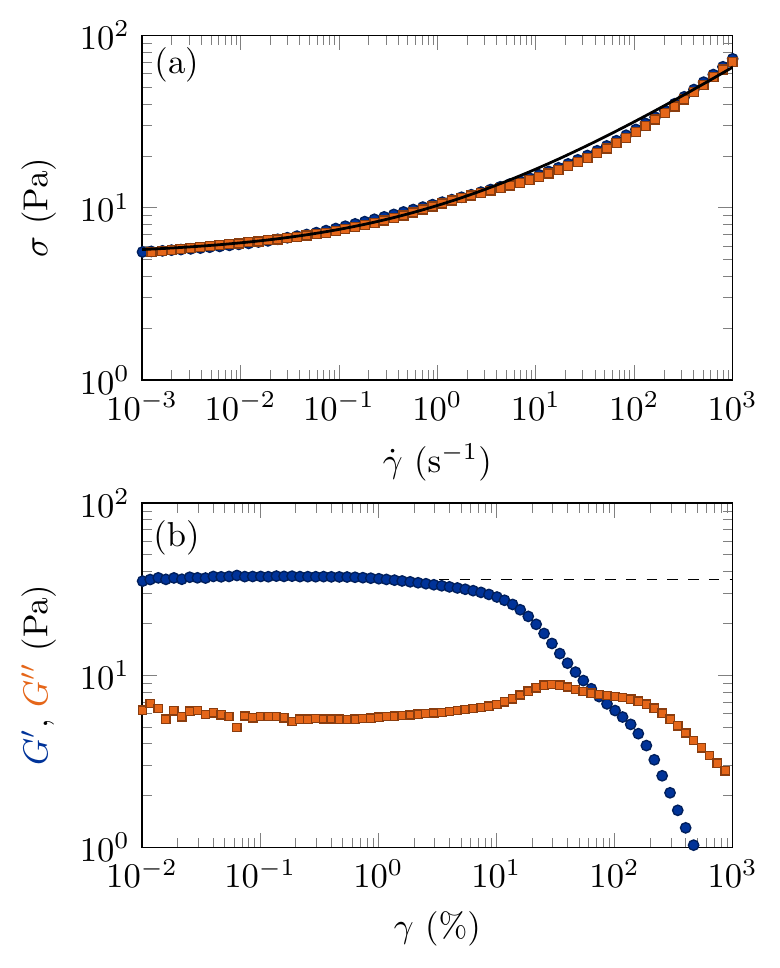}
    \caption{Rheology of the Carbopol microgel. (a) Flow curve. Circles, decreasing shear rates, and squares, increasing shear rates, performed right after the blue circles. The black line represents the best fit to a Herschel-Bulkley law. (b) Oscillatory rheology: amplitude sweep starting from low strain amplitude at $f = 1$~Hz. Circles represent the storage modulus $G'$ while squares show the loss modulus $G''$. The black dashed line represents the average value of the storage modulus in the elastic plateau, $G' = 36$~Pa.}
    \label{fig:app:rheology}
\end{figure}

We characterise the rheology the Carbopol microgel using a standard rotational rheometer (MCR 302) working with a cone-plate geometry fitted with sandpaper discs (grit P1500) to suppress wall slip. Before every test, we apply a pre-shear step at $\dot\gamma = 1500$~s$^{-1}$ for $30$~s and a rest step at $\sigma = 0$~Pa for $20$~s. We perform two consecutive flow curves for decreasing and increasing shear rates $\dot\gamma$ between $0.001$~s$^{-1}$ and $1500$~s$^{-1}$, choosing $5$~s steps and $10$ points per decade. Amplitude sweeps are conducted at a frequency of $1$~Hz for increasing shear strains $\gamma$ between $0.01\%$ to $1000~\%$, and we choose to acquire $15$ points per decade and average over $10$ oscillation cycles. Our data are presented in Figure~\ref{fig:app:rheology}.

Figure~\ref{fig:app:rheology}(a) shows the flow curves of the fluid. The data fit to a Herschel-Bulkley law, $\sigma = \sigma_{\rm Y} + K \dot\gamma^n$ is fair and yields $n = 0.36$, $K = 5.0$~Pa.s$^n$, and $\sigma_{\rm Y} = 5.3$~Pa. The two consecutive flow curves superpose well, meaning that fluid thixotropy is negligible. In Figure~\ref{fig:app:rheology}(b), we identify the linear modulus of the Carbopol $G$ with the storage part of the elastic modulus $G'$ in the linear visco-elastic plateau for which $G' \gg G''$; this plateau spans from $\gamma \geq 0.01\%$ to $\gamma = 10\%$. We obtain $G = 36.0$~Pa. We also notice that the storage modulus is rather insensitive to the applied frequency $f$ in the range accessible to the rheometer, 0.1~Hz to 10~Hz (data not shown).

\section{Contributions to damping of bubble oscillations}
\label{sec:app:damping}

\begin{figure}
    \centering
    \includegraphics{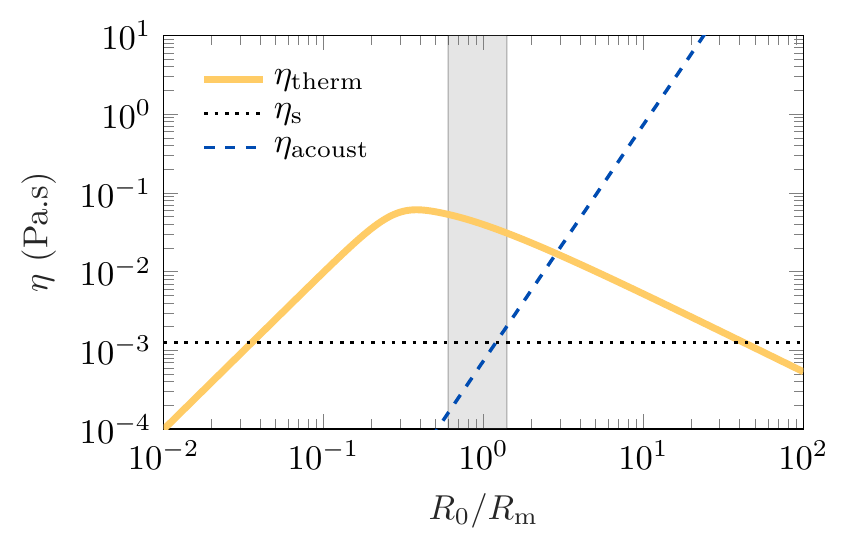}
    \caption{Thermal, viscous and acoustic damping under linear bubble oscillation plotted as effective viscosities for an applied frequency $f= 22.5$~kHz. The thermal and acoustic contributions are directly plotted from Equation~\eqref{eq:damping}, while the value of $\eta_{\rm s}$ has been fitted to the resonance curve in Section~\ref{sec:results_resonancecurve}. The greyed out region is our usual operating range.}
    \label{fig:app:dissipations}
\end{figure}

We compare in Figure~\ref{fig:app:dissipations} the relative magnitude of the three contributions to dissipation detailed in Equation~\eqref{eq:damping}. We respectively note:

\begin{align}
    &\eta_{\rm therm} = \frac{3 p_0 \kappa'}{8 \pi f}
    &\eta_{\rm acoust} = \frac{\pi^2 \rho f^2 R_0^3}{4 c}  
\end{align}
and we plot $\eta_{\rm therm}$, $\eta_{\rm acoust}$ and $\eta_{\rm s}$ for a solvent viscosity $\eta_{\rm s} = 1.3$~mPa.s deduced from Section~\ref{sec:results_resonancecurve}. While acoustic damping is smaller than the viscous term in our operating range, thermal damping dominates them both and is up to $50$ times higher than the viscous contribution. As we fit $\eta_{\rm s}$ by subtracting thermal dissipation $\eta_{\rm therm}$ and acoustic dissipation $\eta_{\rm acoust}$ from the total damping term $\beta$ in the resonance curves, the solvent viscosity $\eta_{\rm s}$ fluctuates greatly for relatively small relative changes in $\eta_{\rm therm}$ and, to a lesser extent, in $\eta_{\rm acoust}$. Obtaining a reliable value of the viscosity $\eta_{\rm s}$ then necessitates very high-quality resonance curve data and precise values of all the physical quantities present in thermal and acoustic damping, which are: $p_0$, $R_0$, $\rho$, $c$ and $D$ through the Péclet number in $\kappa'$. \add{Uncertainties on both $\eta_{\rm therm}$ and $\eta_{\rm acoust}$ have been estimated in ESI Section 1. The lack of precise measurements on the thermal diffusion coefficient $D$ results in a significant uncertainty on $\eta_{\rm therm}$, of the same order as $\eta_{\rm s}$, while the uncertainty on $\eta_{\rm acoust}$ remains negligible}.

\section{\add{Threshold for shape oscillations in soft materials}}
\label{sec:app:critshapeamp}

\add{We recall here the predictions of Ref.~\citenum{Murakami2020}, who derived the critical bubble oscillation amplitude above which shape oscillations may be observed in a neo-Hookean, Kelvin-Voigt viscoelastic solid. Defining intermediate quantities:}

\begin{align}
\lambda_1 &= 4(k - 1)(k + 1)(k + 2) \frac{\Gamma}{4 \pi^2 f^2 \rho R_0^3} \\
\lambda_2 &= 2 (k + 2) (2 k + 1) \frac{\eta_{\rm s}}{2 \pi f \rho R_0^2} \\
\lambda_3 &= 4 (k + 1) \frac{G}{4 \pi^2 f^2 \rho R_0^2} \\
\lambda_4 &= 12 k (k+2) \frac{\eta_{\rm s}}{2 \pi f\rho R_0^2}
\end{align}

\begin{widetext}
\add{The critical amplitude $\zeta_{{\rm c}, k}$ above which a shape mode $k$ develops may be expressed as:}
\begin{equation}
    \zeta_{{\rm c},k } ^2 = \frac{\left [ (\lambda_1 - 1)  + \lambda_3 \left (4 + 4k/3 + k^2/3 \right ) \right  ]^2 + 4 \lambda_2^2 }{\left [ (2k + 1) - 3/2 \lambda_1 + 2\lambda_2^2 - \lambda_3 \left (18 + 19 k/3 + k^2/3 \right )  \right ]^2 + \lambda_4^2}
    \label{eq:critical_amplitude}
\end{equation}
\end{widetext}

\add{Since our experiments show that bubble oscillations up to the shape oscillation threshold are linear in the time domain, we may combine Equations~\eqref{eq:resonance_curve}  and~\eqref{eq:critical_amplitude} to derive explicitly the critical pressure $p_{{\rm c},k}$ for all modes $k$. One surprising consequence of Equation~\eqref{eq:critical_amplitude} is that, despite modelling the three-dimensional growth of spherical harmonics $Y_k^m$ generally defined by two shape modes $k$ and $m$, the critical pressure of the model is independent of $m$.}

\add{Close to $R_0 = R_{\rm m}$, the critical pressure $p_{{\rm c},k}$ reaches a minimum for all modes $k$ because it corresponds to the resonance condition of spherical oscillations. In addition, shape modes have a natural oscillation frequency, given by:} 
\begin{widetext}
\begin{equation}
    \label{eq:lambshapemodes}
    \pi^2 f^2 = \frac{G }{\rho R_0^2} (k + 1 ) \left [ 4 + k + \frac{k (k+1)}{3} \right ] + \frac{\Gamma }{\rho R_0^3}(k+1)(k-1)(k+2) \,.
\end{equation}
\end{widetext} 

\add{When $f$ is imposed, Equation~\ref{eq:lambshapemodes} defines a radius at which a given shape mode $k$ resonates, corresponding to the minima of the coloured tongues in Figure~\ref{fig:pressure_instability}. In some particular cases (here, for $k = 5$ and $6$), both the spherical mode and the shape mode resonate around $R_{\rm m}$, resulting in particularly low critical pressures $p_{{\rm c},5}$ and $p_{{\rm c},6}$, as observed in Figure~\ref{fig:pressure_instability} and in the experiments.}

\bibliographystyle{tex/rsc}
\bibliography{biblio}
\end{document}